# Model-based controller assisted domain randomization for transient vibration suppression of nonlinear powertrain system with parametric uncertainty

Heisei Yonezawa[#1], Ansei Yonezawa[#2], Itsuro Kajiwara [#3]

#1(Corresponding author)
Division of Mechanical and Aerospace Engineering, Hokkaido University
N13, W8, Kita-ku, Sapporo, Hokkaido 060-8628, Japan
Phone +81-80-4039-9406
E-mail: yonezawah@eng.hokudai.ac.jp

#2
Department of Mechanical Engineering, Kyushu University
744 Motooka, Nishi-ku, Fukuoka 819-0395, Japan
Phone +81- 90-2893-3160
E-mail: ayonezawa@mech.kyushu-u.ac.jp

#3
Division of Mechanical and Aerospace Engineering, Hokkaido University
N13, W8, Kita-ku, Sapporo, Hokkaido 060-8628, Japan
Phone +81-11-706-6390, Fax. +81-11-706-6390
E-mail: ikajiwara@eng.hokudai.ac.jp

**Abstract**

Complex mechanical systems such as vehicle powertrains are inherently subject to multiple nonlinearities and uncertainties arising from parametric variations. Modeling errors are therefore unavoidable, making the transfer of control systems from simulation to real-world systems a critical challenge. Traditional robust controls have limitations in handling certain types of nonlinearities and uncertainties, requiring a more practical approach capable of comprehensively compensating for these various constraints. This study proposes a new robust control approach using the framework of deep reinforcement learning (DRL). The key strategy lies in the synergy among domain randomization-based DRL, long short-term memory (LSTM)-based actor and critic networks, and model-based control (MBC). The problem setup is modeled via the latent Markov decision process (LMDP), a set of vanilla MDPs, for a controlled system subject to uncertainties and nonlinearities. In LMDP, the dynamics of an environment simulator is randomized during training to improve the robustness of the control system to real testing environments. The randomization increases training difficulties as well as conservativeness of the resultant control system; therefore, progress is assisted by concurrent use of a model-based controller based on a physics-based system model. Compared to traditional DRL-based controls, the proposed approach is smarter in that we can achieve a high level of generalization ability with a more compact neural network architecture and a smaller amount of training data. The controller is verified via practical application to active damping for a complex powertrain system with nonlinearities and parametric variations. Comparative tests demonstrate the high robustness of the proposed approach.

**Keywords**: Active vibration control, Domain randomization, Robust control, Model-based control, Powertrain, Deep reinforcement learning

## 1. Introduction

In recent years, the performance requirements for industrial mechanical systems, such as automotive powertrain systems [1][2] and robotic control [3][4], have become increasingly sophisticated, leading to a continuous increase in system complexity. Mechanical systems are inherently subject to nonlinearities [5], communication delays, and uncertainties arising from parametric variations [6]. Therefore, control system design must rigorously account for these factors to ensure stability and performance.

Since the formalization of modern control theory by Rudolf Kalman [7], the superior performance of model-based control systems, which rely on system models (e.g., linear state-space equations), has been widely recognized. Inspired by this paradigm, control system design has traditionally focused on achieving high-precision modeling of the controlled system by leveraging governing equations and system identification techniques. However, as industrial systems continue to grow in complexity and exhibit increasing nonlinearities, naïve model-based control approaches are reaching their fundamental limitations. Modeling and calibration errors are inherently unavoidable, making the transfer of control strategies from simulation to real-world systems a critical challenge—commonly referred to as the *sim-to-real gap problem* [8]. Mitigating performance degradation in the presence of system variations and nonlinearities remains a top-priority challenge in both industry and academia and continues to be an unresolved issue. However, traditional robust controls such as $H_\infty$ control [9][10] have inherent limitations in handling certain types of nonlinearities and uncertainties. Industrial systems, such as automotive powertrains, are typically affected by multiple nonlinear characteristics and uncertainties simultaneously. Therefore, more practical control systems are needed capable of comprehensively compensating for these various constraints.

In recent years, the significant advancement in computational power and theoretical breakthroughs in machine learning have brought increasing attention to the integration of artificial intelligence (AI) into control system design [11]. Among these advancements, reinforcement learning (RL) has achieved remarkable success in solving complex, nonlinear, large-scale, and high-dimensional control problems [12][13]. In RL, an *agent*—the decision-making entity—learns to perform complex tasks through trial and error by interacting with the environment. The key advantage of RL lies in its ability to autonomously collect training data and learn without requiring an explicit transition model of the environment's dynamics [14][15].

Notably, deep reinforcement learning (DRL) [16][17], which leverages deep learning techniques, has gained prominence in recent years. A major breakthrough in DRL was the successful approximation of value functions using deep, nonlinear neural networks, significantly enhancing the scalability and generalization capability of RL-based control systems. The proposition of DRL has led to significant advancements, particularly in the extension to continuous action spaces. Notable examples include Deep Deterministic Policy Gradient (DDPG) [18][19] and its improved variant, Twin Delayed Deep Deterministic Policy Gradient (TD3) [20], both of which exemplify the class of policy gradient optimization methods. More recently, Proximal Policy Optimization (PPO) [21] has demonstrated remarkable learning capabilities, underpinning the success of the large language model *ChatGPT*, which has had a profound global impact [22][23]. DRL also has shown the powerful capabilities for controlling complex mechanical systems such as a rotary flexible link manipulator [24], rotating machinery [25], a magnetorheological elastomer vibration absorber [26], and vibration control oriented to flexible solar panels of spacecraft and satellites [27]. The application to multi-story shear building structures showed that DRL outperforms traditional model-based control in partially observed settings [28]. Another previous study demonstrated that integration of DDPG with an iterative gradient-based state feedback control algorithm can accelerate the learning process [29]. For driving a robot gearshift manipulator, a combined approach of active disturbance rejection control and DDPG was proposed [30]. RL inherently relies on trial-and-error exploration, where training data is sampled through stochastic exploratory actions, often resulting in extreme or misguided policies. Consequently, training an RL agent directly in a real-world environment poses significant safety risks [8]. Moreover, collecting the vast amount of experimental data required to acquire an effective policy entails substantial computational costs and time [31]. To alleviate these challenges, a common approach is to train RL agents in simulated environments, where safety concerns are eliminated, and an unlimited amount of training data can be generated [8]. However, the sim-to-real gap is a well-recognized issue not only in control theory but also in RL. Due to modeling and calibration errors, an agent trained to perform a task successfully in simulation may fail to exhibit the same level of performance when deployed in the real world [8]. It should be remarked that previous naïve applications of RL to active vibration controls overlook the issue of sim-to-real gaps.

A practical approach that has been increasingly adopted in the field of robotics to address the sim-to-real gap problem is *domain randomization* (DR) [8,32,33]. While this concept is simple and intuitive, it has proven to be highly effective. To enhance the generalization ability of an RL agent to real-world environments, it is preferable to train the agent under a diverse set of various environments rather than under a single fixed environment. Inspired by this insight, the dynamics of an environment simulator are randomly sampled at the beginning of each training episode in DR. By optimizing the value function across a set of different Markov decision processes (MDPs), the agent can improve generalization and robustness to real-world conditions. One study in particular provided a theoretical analysis on why domain randomization can succeed [33]. From the viewpoint of the successful applications, there are many examples including deformable object manipulation [34], object pushing tasks with 7-DOF robotic arm control [8], locomotion control of bipedal robots [35], imitation learning for real-world animals with agile locomotion skills [36], and humanoid robots [37][38]. Other studies have explored further applications in drone racing [39] and path-following control for an autonomous surface vehicle [31].

However, there remain some technical shortcomings and gaps in previous studies on domain randomization-based techniques from the viewpoint of practical applications as follows:

1. While domain randomization has been increasingly utilized in the field of robotics, its application to a broader range of mechanical systems—such as automotive systems—for positioning and vibration control remains largely unexplored.

2. Since a domain randomized environment is constantly changing due to random sampling, the difficulty of training is increased, hindering the successful learning progress [34]. Additionally, the trained policy is prone to be too conservative [40–42].

Most previous studies on domain randomization do not address these issues. If an enormous amount of training data (episodes) and a large scale of a nonlinear function approximator are available, we may acquire a nearly optimal policy with domain randomization. Nevertheless, such a complex setting not only enlarges the effort, cost, and training time in development, but also raises the risk of overfitting in deep neural networks.

This study focuses on a new controller design to address these problems. It may be questioned whether a control system that relies on a large amount of training data and a large nonlinear function approximator to accomplish its target task can really be called smart. In other words, the central question posed in our study is: *What defines a more intelligent controller*? This study contends that a smarter control system is one that achieves the same level of learning performance (i.e., generalization ability) with *a more compact neural network architecture and a smaller amount of training data*.

Inspired by the above definition, this study proposes an improvement method to overcome the shortcomings of vanilla domain randomization-based control systems. The proposed method is based on the idea that the introduction of a model-based controller (MBC) derived from a nominal model can assist the progress of domain randomization training. The approach is outlined in Fig. 1. The effectiveness of integrating MBC and RL has also been analyzed in recent studies [43], where an advanced model-data hybrid driven control scheme was proposed to address uncertain hydraulic Euler-Lagrange systems. Our present study leverages the similar idea of such integration to enhance the properties of domain randomization. We formulate modeling via the latent Markov decision process (LMDP) for a controlled system subject to uncertainties and nonlinearities. In LMDP, a robust policy is realized by randomizing the dynamics of an environment simulator during training, extending the generalization ability to real testing environments. In the proposed approach, a DRL agent trained by domain randomization compensates for modeling errors (e.g., nonlinearity and parametric uncertainty) while a model-based controller undertakes the base for the control action from scratch, making the training process easier.

In summary, the contributions and technical novelties of this study are as follows:

1. **Integration of MBC with domain randomization in a DRL framework:**
   The essential novelty of this study lies in strengthening domain randomization through the assistance of an MBC designed from a nominal system model. Our novel framework realizes a strong synergy between physical model-based control and data-driven learning, which overcomes many challenges of controlling mechanical systems. By providing baseline control performance from the beginning, MBC significantly improves training efficiency, reduces unnecessary exploration, and mitigates the conservativeness of the resulting policy. This synergy allows the RL agent to achieve robust generalization with fewer training data and smaller network structures.

2. **Memory-augmented policy representation:**
   We employ LSTM-based actor and critic networks, which store past experiences in hidden states and thereby enable the policy to infer the dynamics of randomly varied environments. This enhances the adaptability of the control system under uncertainties.
3. **Formulation as LMDP:**
   From a theoretical perspective, we model the control problem with uncertainties and nonlinearities as an LMDP. We demonstrate that even when the learning environment is extended to include MBC, the Markov property—the theoretical foundation of RL—remains satisfied. This provides a more general theoretical basis for applying the proposed algorithm to a broader class of nonlinear systems and state-space-represented model-based controllers.
4. **Application to nonlinear powertrain vibration control:**
   The proposed approach is validated through numerical simulations for active damping of a vehicle powertrain subject to strong nonlinearities and parametric variations. To the best of our knowledge, this is the first application of domain randomization to complex active vibration control. Comparative validations confirm the superior robustness of the proposed method over existing approaches.

The remainder of this paper is organized as follows. Section 2 describes the problem formulation. Section 3 outlines the standard reinforcement learning setup. Section 4 details the proposed approach, i.e., MBC assisted domain randomization training (MBCA-DRT) strategy. The proposed scheme is extended to the continuous action RL algorithm: DDPG. Section 5 applies the improved robust DDPG to active vibration control of an uncertain powertrain system. The robustness of the proposed approach is verified via numerical examples using the powertrain model. Finally, Section 6 concludes the paper.

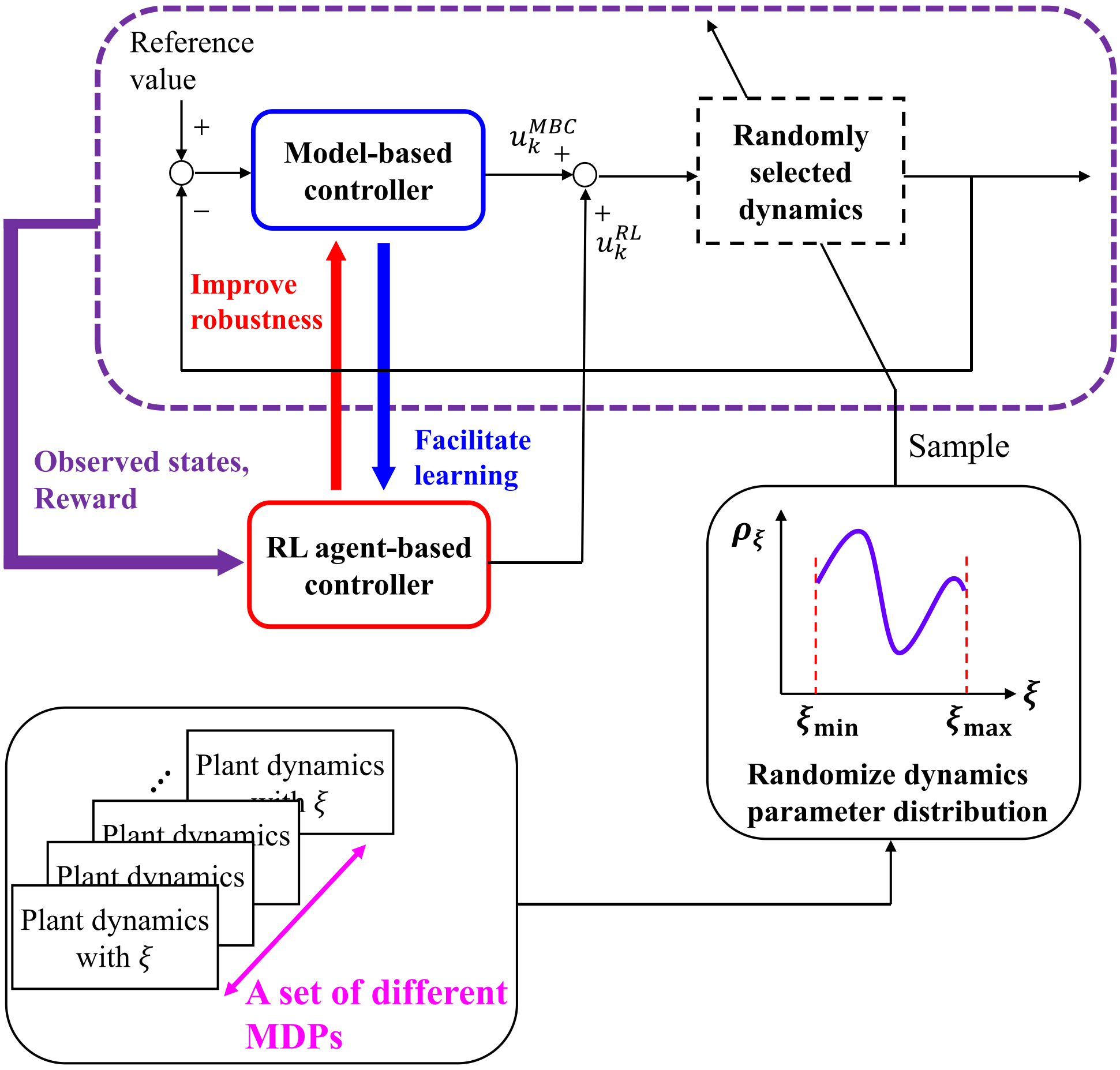

Fig. 1 Outline of MBC assisted domain randomization.

## 2. Problem formulation: nonlinear adaptive optimal control problem

Consider a discrete-time (DT), nonlinear dynamical system defined by

$$x_{k+1} = f(x_k, u_k, w_k;\ \xi) \quad (1)$$

$$y_k = h(x_k;\ \xi) \quad (2)$$

where $x_k \in \mathbb{R}^n$, $u_k \in \mathbb{R}^m$, and $w_k \in \mathbb{R}^l$ are the state, the control input, and the disturbance at discrete time $k$, respectively. The controlled output is denoted by $y_k \in \mathbb{R}^o$. Here, $f(\cdot): \mathbb{R}^n \times \mathbb{R}^m \times \mathbb{R}^l \to \mathbb{R}^n$ is the uncertain nonlinear function that represents the transition dynamics of the system. Also, $h(\cdot): \mathbb{R}^n \to \mathbb{R}^o$ is the uncertain function that defines the system output. The symbol $\xi$ corresponds to a set of system parameters that may be subject to uncertainties such as parameters variations and nonlinearities. The dynamics of the simulation is parametrized by $\xi$ during domain randomization training. The goal of this control problem is to determine the control input $u_k$ at each time step so that the tracking error converges to zero: $\lim_{k\to\infty} \|e_k\| = 0$, where $e_k := y_k^r - y_k$. Here, $y_k^r \in \mathbb{R}^o$ represents a reference value. Note that $w_k$ and $y_k^r$ are independent from past states and control inputs (i.e., they are exogenous disturbances with known or random distributions). Associated with this system, the adaptive optimal control problem with a finite time-horizon is defined based on the immediate cost function $c: \mathbb{R}^n \times \mathbb{R}^m \rightarrow \mathbb{R}$:

$$\min_{\{u_1, u_2, \cdots u_T\} \in U} \sum_{k=0}^{T} c(x_k, u_k) \tag{3}$$

subject to $x_{k+1} = f(x_k, u_k, w_k;\ \xi)\ \forall k \in [0, T]$

where $c(0,0) = 0$ and $c(x_k, u_k) \rightarrow \infty$ as $\|(x_k, u_k)\| \rightarrow \infty$. In this paper, the quadratic cost function is used:

$$c(x_k, u_k) \simeq e_k{}^T Q e_k + u_k{}^T R u_k \tag{4}$$

where $Q \in \mathbb{R}^{o \times o}$ and $R \in \mathbb{R}^{m \times m}$ are positive semi-definite and positive definite symmetric matrices, respectively.

It is not possible to directly apply the conventional model-based linear controller to the optimization problem (3) because the system dynamics (1) and (2) are unknown due to the model uncertainty and nonlinearity. Nevertheless, in most industrial systems, partially known system dynamics can be leveraged to derive the linearized nominal model, which should be useful to boost the efficiency of RL agent training. The following assumptions are introduced into the optimization problem.

***Assumption 1***: The linearized approximate model (i.e., nominal model) of the nonlinear system in Eqs. (1) and (2) is known:

$$x_{k+1} \simeq A x_k + B_1 w_k + B_2 u_k \tag{5}$$

$$y_k \simeq C x_k \tag{6}$$

where $A \in \mathbb{R}^{n \times n}$, $B_1 \in \mathbb{R}^{n \times l}$, $B_2 \in \mathbb{R}^{n \times m}$, and $C \in \mathbb{R}^{o \times n}$. These system matrices can be obtained from various approaches such as first principle-based modeling, system identification, and data-driven approaches.

***Assumption 2***: For the above linear approximate model, we can design a model-based linear controller in the form of the following discrete state-space representation, generating the model-based control input $u_k^{MBC} \in \mathbb{R}^m$. This is an approximate solution for the optimization problem (3).

$$x_{k+1}^c = A_c x_k^c + B_c e_k \tag{7}$$

$$u_k^{MBC} = C_c x_k^c + D_c e_k \tag{8}$$

where $A_c \in \mathbb{R}^{n \times n}$, $B_c \in \mathbb{R}^{n \times o}$, $C_c \in \mathbb{R}^{m \times n}$, and $D_c \in \mathbb{R}^{m \times o}$. Here, $x_k^c \in \mathbb{R}^n$ is an internal state vector of the MBC.

It should be noted that only the model-based fixed controller itself fails to appropriately control the real system (1) and (2) due to the modeling error $\Delta x = f(x_k, u_k, w_k, \xi) - (A x_k + B_1 w_k + B_2 u_k)$ such as parametric uncertainty and nonlinearity neglected in the approximated system (5) and (6). Such nonlinearity negatively affects the control system's performance, causing serious degradation and instability.

RL techniques are capable of dealing with system nonlinearities by training a control policy $\pi$ such that $u_k = \pi(x_k)$ provides approximately optimal solutions to the problem (3). In most RL algorithms, Eqs. (1)-(4) can be equivalently formulated as a DT Markov decision process (MDP) [14][44]; however, the sim-to-real gap is not explicitly addressed.

## 3. Deep reinforcement learning

### 3.1. Markov decision process

This chapter follows the RL framework based on descriptions in literatures [18][19]. Following the convention adopted in most RL literatures, the time step is denoted by $t$, which corresponds to $k$ in Chapter 2.

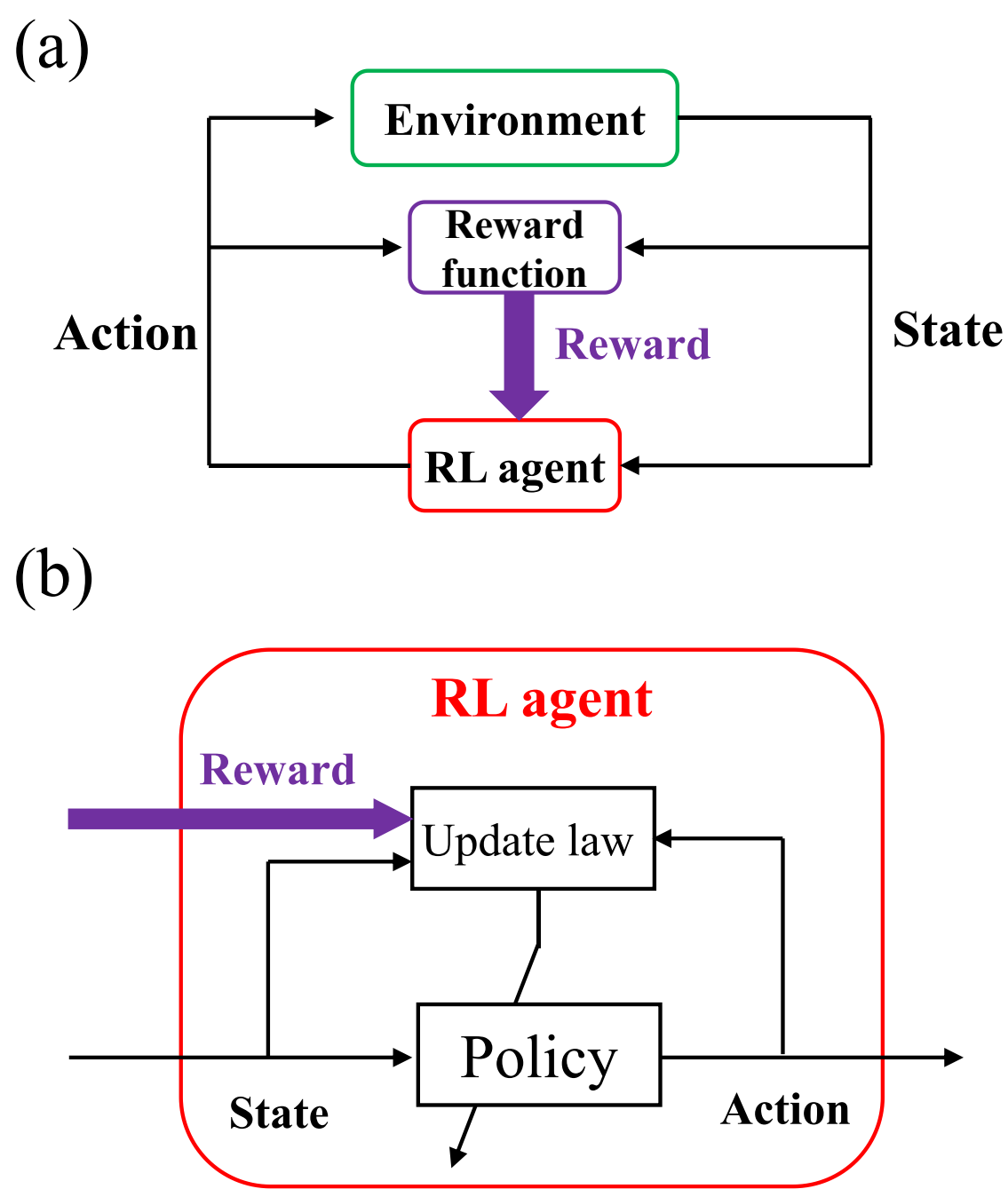

Fig. 2 Setup of standard reinforcement learning.

We consider a standard reinforcement learning setup in which an agent acts in a stochastic environment $E$ by sequentially selecting actions over a sequence of discrete time steps $t$, in order to maximize a cumulative long-term reward, as shown in Fig. 2. We model the problem as an MDP which is composed: a state space $\mathcal{S} = \mathbb{R}^n$, an action space $\mathcal{A} = \mathbb{R}^m$, an initial state distribution with density $p_1(s_1)$, a stationary transition dynamics distribution with conditional density $p(s_{t+1} \mid s_t, a_t)$. The Markov property is satisfied, meaning $p(s_{t+1} \mid s_1, a_1, \ldots, s_t, a_t) = p(s_{t+1} \mid s_t, a_t)$, for any trajectory $s_1, a_1, s_2, a_2, \ldots, s_T, a_T$ in state-action space. At each time step $t$, the agent receives a state $s_t$, takes an action $a_t$ and receives a scalar-valued reward $r_t$. The actions are real-valued $a_t \in \mathbb{R}^m$. A reward function $r(s,a): \mathcal{S} \times \mathcal{A} \rightarrow \mathbb{R}$ is a metric that represents the immediate desirability of a given state and an agent's action of the environment. To evaluate the quality of an agent's actions, it is provided as a scalar value to the agent from the environment at every step. A policy is used to select actions in the MDP. In general, the policy is stochastic and defined by mapping from states to a probability distribution over the actions $\pi_\theta: \mathcal{S} \rightarrow \mathcal{P}(\mathcal{A})$, where $\mathcal{P}(\mathcal{A})$ is a set of probability measures on $\mathcal{A}$ and $\theta \in \mathbb{R}^{d1}$ is a parameter vector of $d1$ dimensions. $\pi_\theta(a_t \mid s_t)$ is the conditional probability density at $a_t$ treated as the policy. The policy is used by the agent to interact with the MDP to generate a trajectory of states, actions and rewards, $\tau_{1:T} = s_1, a_1, r_1 \dots, s_T, a_T, r_T$ over $\mathcal{S} \times \mathcal{A} \times \mathbb{R}$. The return $r_t^\gamma$ is the sum of discounted future reward from time-step $t$ onwards, $r_t^\gamma = \sum_{i=t}^{T} \gamma^{i-t} r(s_i, a_i)$ where $0 < \gamma < 1$ is a discounting factor.

Value function is used as the objective function in many RL algorithms. It is defined to be the expected total discounted reward, starting from state $s$ and thereafter following policy $\pi$:

$$V^\pi(s) = \mathbb{E}\left[r_1^\gamma \mid S_1 = s; \pi\right]. \tag{9}$$

Similarly, action-value function is defined to be the expected return after taking an action $a$ in state $s$ and thereafter following policy $\pi$:

$$Q^\pi(s,a) = \mathbb{E}\left[r_1^\gamma \mid S_1 = s, A_1 = a; \pi\right]. \tag{10}$$

The goal of the agent in RL is to learn a policy which maximizes the expected cumulative discounted reward from the start state, defined by the performance objective $J(\pi) = \mathbb{E}\left[r_1^\gamma \mid \pi\right]$.

RL approaches make use of the recursive relationship known as the Bellman equation, which is a fundamental relationship between the value of a state-action pair $(s_t, a_t)$ and the value of the subsequent state-action pair $(s_{t+1}, a_{t+1})$:

$$Q^\pi(s_t, a_t) = \mathbb{E}_{r_t, s_{t+1} \sim E}\left[r(s_t, a_t) + \gamma \mathbb{E}_{a_{t+1} \sim \pi}\left[Q^\pi(s_{t+1}, a_{t+1})\right]\right]. \tag{11}$$

For the target deterministic policy $\mu: \mathcal{S} \rightarrow \mathcal{A}$, we can remove the inner expectation:

$$Q^\mu(s_t, a_t) = \mathbb{E}_{r_t, s_{t+1} \sim E}\left[r(s_t, a_t) + \gamma Q^\mu\left(s_{t+1}, \mu(s_{t+1})\right)\right]. \tag{12}$$

We describe the density at state $s'$ after transitioning for $t$ time steps from state $s$ by $p(s \rightarrow s', t, \pi)$. We also describe the (improper) discounted state visitation distribution for a policy $\pi$ by $\rho^\pi(s') := \int_{\mathcal{S}} \sum_{t=1}^{\infty} \gamma^{t-1} p_1(s) p(s \rightarrow s', t, \pi) ds$. The performance objective is then defined as an expectation as:

$$J(\pi_\theta) = \int_{\mathcal{S}} \rho^\pi(s) \int_{\mathcal{A}} \pi_\theta(s,a) r(s,a) da\, ds = \mathbb{E}_{s \sim \rho^\pi, a \sim \pi_\theta}[r(s,a)] \tag{13}$$

where $\mathbb{E}_{s \sim \rho}[\cdot]$ indicates the (improper) expected value with respect to discounted state distribution $\rho(s)$.

### 3.2. Policy gradient method based on actor and critic algorithm

For control problems with continuous action space, policy gradient methods are the most powerful class of RL algorithms. The fundamental idea is to iteratively adjust the policy parameters $\theta \in \mathbb{R}^{d1}$ in the direction of the gradient $\nabla_\theta J(\pi_\theta)$ of the performance objective. The policy gradient theorem is the basic result underlying these algorithms [45]:

$$\nabla_\theta J(\pi_\theta) = \int_{\mathcal{S}} \rho^\pi(s) \int_{\mathcal{A}} \nabla_\theta \pi_\theta(a \mid s) Q^\pi(s,a) da\, ds \\ = \mathbb{E}_{s\sim\rho^\pi, a\sim\pi_\theta}[\nabla_\theta \log\, \pi_\theta(a \mid s) Q^\pi(s,a)]. \tag{14}$$

The expectation $\mathbb{E}_{s\sim\rho^\pi, a\sim\pi_\theta}$ can be formed through an empirical average of a sample-based estimation (approximation). However, one problem with this algorithm is how to estimate the action-value function $Q^\pi(s,a)$.

In order to estimate both the policy gradient and $Q^\pi(s,a)$ simultaneously, the *actor-critic* algorithm is widely employed based on the policy gradient theorem [45–48]. The *actor-critic* method relies on two eponymous components. An *actor* tunes the parameters $\theta$ of the policy $\pi_\theta(s)$ by stochastic gradient ascent of Eq. (14). Instead of the unknown true action-value function $Q^\pi(s,a)$ in Eq. (14), an estimation of an action-value function $Q^\phi(s,a)$ is employed with parameter vector $\phi$. The role of a *critic* is to estimate the action-value function $Q^\phi(s,a) \approx Q^\pi(s,a)$ based on an appropriate policy evaluation algorithm such as temporal-difference learning.

A concern with the above architecture is that estimation bias may be introduced by substituting a function approximator $Q^\phi(s,a)$ for the true action-value function $Q^\pi(s,a)$. However, if the function approximator is appropriately selected such that i) $Q^\phi(s,a) = \nabla_\theta \log\, \pi_\theta(a \mid s)^\top \phi$ and ii) the parameters $\phi$ are determined to minimize the mean-squared error $\epsilon^2(w) = \mathbb{E}_{s\sim\rho^\pi, a\sim\pi_\theta}\left[\left(Q^\phi(s,a) - Q^\pi(s,a)\right)^2\right]$, then no bias is induced [45],

$$\nabla_\theta J(\pi_\theta) = \mathbb{E}_{s\sim\rho^\pi, a\sim\pi_\theta}\left[\nabla_\theta \log\, \pi_\theta(a \mid s) Q^\phi(s,a)\right]. \tag{15}$$

In an off-policy setting [49], we can estimate the policy gradient from trajectories sampled from a different stochastic exploration policy $\beta(a \mid s) \neq \pi_\theta(a \mid s)$ using an importance sampling ratio $\pi_\theta(a \mid s)/\beta(a \mid s)$ as:

$$\nabla_\theta J_\beta(\pi_\theta) \approx \int_{\mathcal{S}} \int_{\mathcal{A}} \rho^\beta(s) \nabla_\theta \pi_\theta(a \mid s) Q^\pi(s,a) da\, ds \\ = \mathbb{E}_{s\sim\rho^\beta, a\sim\beta}\left[\frac{\pi_\theta(a \mid s)}{\beta_\theta(a \mid s)} \nabla_\theta \log \pi_\theta(a \mid s)\, Q^\pi(s,a)\right]. \tag{16}$$

### 3.3. Deep deterministic policy gradient (DDPG)

The Deep Q Network (DQN) algorithm [16][17] is a method based on Q-learning that successfully leverages a deep neural network-based function approximator for estimating the action-value function. Furthermore, an extension of DQN to continuous action spaces within the off-policy actor-critic framework has been developed called the deep deterministic policy gradient (DDPG) algorithm [19]. While the policy gradient in DDPG is derived from the deterministic policy gradient (DPG) theorem [18], it employs deep neural networks as function approximators for both the actor policy and the critic value function.

Q-learning [50], a commonly used off-policy algorithm, employs the greedy maximization (or soft maximization) of the estimated action-value function: $\mu(s) = \arg\max_a Q^\mu(s,a)$. We consider a deep neural network function approximator $Q^{\phi}(s, a)$ in place of the true action-value function $Q^{\mu}(s, a)$. The function approximator is parameterized by $\phi \in \mathbb{R}^{d2}$, which is optimized by minimizing the critic loss:

$$L(\phi) = \mathbb{E}_{s_t \sim \rho^{\beta}, a_t \sim \beta, r_t \sim E}\left[\left(Q^{\phi}(s_t, a_t) - Y_t\right)^2\right] \tag{17}$$

where

$$Y_t = r(s_t, a_t) + \gamma Q^{\phi}\big(s_{t+1}, \mu(s_{t+1})\big). \tag{18}$$

While $Y_t$ also depends on $\phi$, this is usually ignored.

Q-learning is not applicable to control problems with continuous action spaces because the greedy policy assumes maximization of the action-value function over a discrete action space, which is computationally impractical in a continuous space. In contrast, the DDPG algorithm optimizes both a policy and an action-value function, approximated by deep neural networks, within the actor-critic framework under the assumption of a deterministic policy $\mu_{\theta}: \mathcal{S} \to \mathcal{A}$ parameterized by $\theta \in \mathbb{R}^{d1}$. This approach is applicable to continuous action spaces. In DDPG, similar to DQN, the critic $Q^{\phi}(s, a)$ is updated by minimizing the mean-squared error based on the Bellman equation. Meanwhile, the actor optimizes the policy function $\mu_{\theta}$, which defines a deterministic mapping from states to actions. Specifically, the actor's parameters $\theta$ are updated in the direction of the gradient of the estimated action-value function with respect to the actor parameters. Since the update direction varies for each visited state, the expectation is taken over the off-policy state distribution $\rho^{\beta}$. Defining a performance objective $J(\mu_{\theta}) = \mathbb{E}\left[r_1^{\gamma} \mid \mu_{\theta}\right]$, applying the chain rule results in the DPG theorem [18]:

$$\begin{aligned}\nabla_{\theta} J(\mu_{\theta}) &\approx \mathbb{E}_{s_t \sim \rho^{\beta}}\left[\nabla_{\theta} Q^{\phi}(s, a)\big|_{s=s_t, a=\mu_{\theta}(s_t)}\right] \\ &= \mathbb{E}_{s_t \sim \rho^{\beta}}\left[\nabla_{a} Q^{\phi}(s, a)\big|_{s=s_t, a=\mu_{\theta}(s_t)} \nabla_{\theta}\mu_{\theta}(s)\Big|_{s=s_t}\right].\end{aligned} \tag{19}$$

To improve the stability and robustness of learning, several techniques have been incorporated into DQN and DDPG [16][17]. One such technique is off-policy mini-batch learning using a replay buffer, which minimizes correlations between samples and enhances optimization efficiency. Additionally, the target networks ${Q_{tn}}^{\phi_{tn}}$ and $\mu_{\theta_{tn}}^{tn}$ with weights $\phi_{tn}$ and $\theta_{tn}$ are introduced for computing $Y_t$ in Eq. (18). The weights $\phi_{tn}$ and $\theta_{tn}$ are updated as: $\phi_{tn} \leftarrow \eta\phi + (1-\eta)\phi_{tn}$ and $\theta_{tn} \leftarrow \eta\theta + (1-\eta)\theta_{tn}$ with $\eta \ll 1$, respectively. By delaying the updates of the target networks, optimization stability is improved.

In summary, the following deterministic off-policy actor-critic algorithm is formulated [18][19]:

$$\delta_t = r_t + \gamma {Q_{tn}}^{\phi_{tn}}\left(s_{t+1}, \mu_{\theta_{tn}}^{tn}(s_{t+1})\right) - Q^{\phi}(s_t, a_t) \tag{20}$$

$$\phi_{t+1} = \phi_t + \alpha_{\phi}\delta_t \nabla_{\phi} Q^{\phi}(s_t, a_t) \tag{21}$$

$$\theta_{t+1} = \theta_t + \alpha_{\theta}\mathbb{E}_{s_t \sim \rho^{\beta}}\left[\nabla_{a} Q^{\phi}(s, a)\big|_{s=s_t, a=\mu_{\theta}(s_t)} \nabla_{\theta}\mu_{\theta}(s)\Big|_{s=s_t}\right]. \tag{22}$$

In continuous action spaces, DDPG employs the following stochastic exploration policy $\beta(\cdot)$ with sampling from a noise process $\mathcal{N}$ to ensure sufficient exploration [19].

$$\beta(s_t) = \mu_\theta(s_t) + \mathcal{N}. \tag{23}$$

## 4. Proposed approach

### 4.1. Latent Markov decision process with domain randomization

In this study, the dynamics (e.g., uncertain physical parameters) of the environment used as a simulator are randomized during training to expose the RL agent to a diverse range of environments. Through this process, the RL agent acquires the capability to adapt to mismatches between the simulator and real worlds, thereby enhancing its generalization and robustness when deployed in a real environment. This approach, known as *domain randomization*, has recently gained significant attention especially in the field of robotics, where precise modeling is often impractical for real-world applications [8].

To enhance robustness against uncertainties, it is preferable to train the agent to achieve effective control under various simulator dynamics rather than in a single specific environment. Hence, the simulator is modeled as a set of MDPs parameterized by unknown dynamics parameters $\xi$. In domain randomization, a set of uncertain parameters $\xi$ is randomly changed at the beginning of each episode, according to the distribution $\rho_\xi$ [8]. The parameter set $\xi$ remains fixed throughout an episode and is resampled at the beginning of the next episode [8][32]. Consequently, the control problem can be regarded as a set of different MDPs, where each MDP is associated with its own latent variables, i.e., a set of the parameters $\xi$. Such a control problem can be formulated as a latent Markov decision process (LMDP) where the latent variable can select different MDPs [32,33,51]. The latent variables are introduced to capture variations of a set of the parameters $\xi$, and each single trajectory $\tau_{1:T}$ under a randomly sampled environment can be regarded as an MDP with finite horizon $T$.

Formally, we consider a set of the parameters $\xi$ to parameterize the dynamics of the simulator (environment) $p(s_{t+1} \mid s_t, a_t, \xi)$ [8]. $\mathcal{M}$ is defined as a set of MDPs with finite horizon $T$ under different sets of the dynamics parameters $\xi$. Let the LMDP be denoted as $(\mathcal{M}, \rho_\xi)$, where $\rho_\xi$ denotes the distribution of $\xi$ (i.e., the latent variables) over $\mathcal{M}$ [32][33]. Under these assumptions, we can modify the objective function such that the expected return is maximized across the distribution of dynamics models [8,32,33]:

$$J(\pi_\theta) = \mathbb{E}_{\xi\sim\rho_\xi}\left[\mathbb{E}_{\tau\sim p(\tau\mid\pi_\theta,\xi)}\left[\sum_{i=1}^{T} \gamma^{i-1} r(s_i, a_i)\right] \mid s_1, a_1; \pi_\theta\right] \tag{24}$$

where the likelihood of a trajectory $\tau = (s_1, a_1, s_2, \dots, a_{T-1}, s_T)$ is denoted by $p(\tau \mid \pi_\theta, \xi)$ under the policy $\pi_\theta$.

In the simulator modeled as LMDP, each MDP is randomly selected from a set of MDPs according to the distribution $\rho_\xi$ at the beginning of each episode. The training process is to find an optimal history-dependent policy for LMDP [32][33].

$$\pi_\theta^* = \arg\max_{\pi_\theta} J(\pi_\theta). \tag{25}$$

Nevertheless, one serious issue remains: it becomes more difficult for the RL agent to learn an effective control policy from scratch because the simulator, i.e., environment, is subject to unceasing random sampling in domain randomization. Furthermore, the resulting policy is prone to be more conservative especially when the simulator has many aspects of randomized dynamics [40–42].

### 4.2. Model-based controller assisted domain randomization training (MBCA-DRT)

This study proposes a new idea in which domain randomization improves the generalization performance of an RL agent to real testing environments while the training is efficiently assisted by a MBC. The proposed approach is outlined in **Algorithm 1**. Both the RL agent and the MBC are complementary to each other. From the viewpoint of the RL agent, the MBC eliminates the need for many meaningless trials that were due to immature policy search during initial iterations. In other words, MBC supports the RL agent by stabilizing the balance between exploration and exploitation, boosting the learning convergence. Because MBC is based on a nominal system model, it can form the base of the control action from the bottom. Meanwhile, from the viewpoint of MBC, the role of the RL agent is to ensure robustness by compensating for nonlinearity and uncertainty due to sim-to-real gaps. Because the RL agent does not need to undertake the entire control action from scratch, its learning process will be easier. Therefore, only the minimum data and simple network architecture are sufficient for training. The proposed control input is defined as the linear combination of an RL agent's policy $u_k^{RL} \leftarrow \pi_\theta$ and a MBC $u_k^{MBC}$ as follows:

$$u_k = u_k^{RL} + u_k^{MBC} \tag{26}$$

$$u_k^{RL} \leftarrow \pi_\theta \tag{27}$$

The property that the transition probability dynamics to the next state depend only on the current state and the current action, without being influenced by past states, is called the Markov property [15]. When applying RL, it is desirable for the environment dynamics to follow MDP [52].

To apply domain randomization, we formulate the simulator as an LMDP; a set of MDPs with tunable latent variables $\xi$ and finite horizon $T$, where the latent variable can select different MDPs. Each single process can be regarded as an MDP with finite horizon $T$. For the hybrid approach in Eqs. (26), we need to consider whether each single environment randomly sampled at the beginning of each episode has a Markov property or not. As a general principle, the following proposition can be established.

***Proposition***: Suppose that ***Assumptions 1 and 2*** hold. The closed-loop system is configured with the nonlinear system in Eqs. (1) and (2) and the controller in Eqs. (7) and (8), corresponding to each variable $\xi$ randomly sampled. Let this closed-loop system be denoted as environment *E* randomly sampled at the beginning of each episode. Let the state of environment *E* be $\mathcal{X}_k^{env} \triangleq \left[{x_k}^T \quad {x_k^c}^T\right]^T$. Then, the state transition of environment *E* satisfies the Markov property; therefore, LMDP can be formulated with Eq. (26).

***Proof***: The proof follows a similar scheme to [52][53]. The parameter set $\xi$ sampled through domain randomization is uniquely determined for each environment *E*. Furthermore, the state $x_{k+1}$ of the controlled system in Eqs. (1) and (2) at time $k+1$ is uniquely determined by the state $x_k$ and the control input $u_k$ at time $k$. Note that $w_k$ and $y_k^r$ are independent from all past states and control inputs, as they are exogenous disturbances. In Eq. (26), $u_k^{MBC}$ is computed with Eq. (8). Therefore, the state transition of environment *E*, which integrates the controlled system and the model-based controller as the closed-loop system, is given as follows.

$$\begin{aligned}
\mathcal{X}_{k+1}^{env} = \begin{bmatrix} x_{k+1} \\ x_{k+1}^c \end{bmatrix} &= \begin{bmatrix} f(x_k, u_k, w_k;\ \xi) \\ A_c x_k^c + B_c e_k \end{bmatrix} = \begin{bmatrix} f\left(x_k, u_k^{RL} + u_k^{MBC}, w_k;\ \xi\right) \\ A_c x_k^c + B_c e_k \end{bmatrix} \\
&= \begin{bmatrix} f\left(x_k, u_k^{RL} + C_c x_k^c + D_c(y_k^r - y_k), w_k;\ \xi\right) \\ A_c x_k^c + B_c(y_k^r - y_k) \end{bmatrix} \\
&= \begin{bmatrix} f\left(x_k, u_k^{RL} + C_c x_k^c + D_c y_k^r - D_c h(x_k;\ \xi), w_k;\ \xi\right) \\ A_c x_k^c + B_c y_k^r - B_c h(x_k;\ \xi) \end{bmatrix} \\
&= \mathcal{F}^{env}\left(\mathcal{X}_k^{env}, u_k^{RL}, w_k, y_k^r;\ \xi\right) \\
\mathcal{F}^{env}(\cdot) &\triangleq \begin{bmatrix} f(*) \\ A_c x_k^c + B_c e_k \end{bmatrix}.
\end{aligned} \tag{28}$$

According to Eq. (28), the state $\mathcal{X}_{k+1}^{env}$ of environment *E* at time $k+1$ is uniquely determined by the state $\mathcal{X}_k^{env}$ and the control input $u_k^{RL}$ at time $k$. Therefore, the state transition dynamics of each environment *E* follows MDP.■

**Algorithm 1:** The MBCA-DRT Algorithm.

Input: Environment $E$ and trajectories pool $\mathcal{T}_{rand}$
Output: Optimal history-dependent policy
Randomly initialize the actor's policy parameters and the critic parameters.
Design a MBC with a nominal system model.
**for** each episode **do**
  $\xi \sim \rho_\xi$ randomly sample dynamics.
  $E \leftarrow \Xi(\xi)$ randomly selected environment.
  Generate rollout $\tau = (s_1, a_1, \cdots, s_T) \sim \pi_\theta + u_{MBC}$ with dynamics $\xi$ and $E$.
  $\mathcal{T}_{rand} \leftarrow \mathcal{T}_{rand} \cup \tau_i$
  $u_{RL} \leftarrow \pi_\theta$ RL control input from the actor policy.
  $u = u_{RL} + u_{MBC}$ combination of the RL policy $u_{RL}$ and the model-based control $u_{MBC}$.
  **for** each gradient step **do**
    Compute the internal memories of the recurrent actor and critic networks.
    Update the critic parameters by minimizing the mean squared error.
    Update the actor policy parameters using the sampled policy gradient.
    $\theta \leftarrow \theta + \alpha_\theta \nabla_\theta J(\pi_\theta)$ with $\mathcal{T}_{rand}$ update.
    $u = u_{RL} + u_{MBC}$ update the hybrid control.
  **end**
**end**

***Remark 1***: The above proposition means that the accumulated practical knowledge in previous research of domain randomization is also applicable to the combined approach in Eq. (26). This is because each environment randomly selected at the beginning of each episode follows the Markov property, even if it includes a MBC. In other words, like vanilla domain randomization, we can formulate LMDP for a set of those closed-loop systems.

### 4.3. Memory-augmented actor and critic based on long short-term memory (LSTM)

The actor and critic networks employ a long short-term memory (LSTM) architecture to provide the agent with a mechanism for inferring the dynamics of an environment [8][32]. The LSTM network is well-suited for effectively handling long-term dependencies in sequential data [54]. Since past state and action information is stored in the internal memories, the agent can improve its generalization ability to varying environments [8].

The inputs of LSTM are the current state $s_t$, the previous hidden state $h_{t-1}$, and the previous cell state $c_{t-1}$. The previous hidden state and previous cell state can effectively store previous dynamics information, which is used to improve the generalization capability. The forget, input, and output gates are used to control what information should be forgotten, retained, and outputted. The three gates are mathematically formulated as follows [54]:

$$f_t = \sigma\left(W_f s_t + R_f h_{t-1} + b_f\right), \tag{29}$$

$$i_t = \sigma(W_i s_t + R_i h_{t-1} + b_i), \tag{30}$$

$$o_t = \sigma(W_o s_t + R_o h_{t-1} + b_o), \tag{31}$$

where $W_*$, $R_*$, and $b_*$ denote the input weight, recurrent weight, and bias, respectively. $\sigma(\cdot)$ represents the activation function. The current cell state $c_t$ is given by

$$c_t = f_t \odot c_{t-1} + i_t \odot g_t \tag{32}$$

where $\odot$ represents the Hadamard product, and $g_t$ is the cell state candidate given by

$$g_t = \sigma\left(W_g s_t + R_g h_{t-1} + b_g\right). \tag{33}$$

The current hidden state is computed as follows:

$$h_t = o_t \odot \tanh\left(c_t\right) \tag{34}$$

With the memory-augmented actor and critic architecture, the information regarding dynamics of the environment (each simulator selected from LMDP) can be encoded as hidden states, assisting the RL agent to infer more accurate environment dynamics [8,32,33].

For the LSTM-based actor and critic networks, we propose to leverage information on the control input $u_{MBC}$ from MBC for training an RL agent. This additional observed information allows for the RL agent to explicitly learn a mechanism by which modeling errors aggravate the nominal controller's performance, which is helpful for the agent to find an effective compensation strategy realizing robustness. This configuration is approximately realized in actual implementation.

### 4.4. DDPG algorithm with MBCA-DRT

In this study, we employ the DDPG algorithm [19] to train an RL agent. In addition, a domain randomization technique is also integrated with DDPG, which is supported by the MBC to boost learning convergence. The proposed algorithm is summarized in **Algorithm 2**. The outline is illustrated in Fig. 3.

**Algorithm 2:** DDPG algorithm with MBCA-DRT.

Randomly initialize critic network $Q^{\phi}(s, a)$ and actor $\mu_{\theta}(s)$ with weights $\phi$ and $\theta$.
Initialize target networks $Q_{tn}{}^{\phi_{tn}}$ and $\mu_{\theta_{tn}}^{tn}$ with weights $\phi_{tn} \leftarrow \phi$ and $\theta_{tn} \leftarrow \theta$.
Initialize replay buffer $R$.
Design a MBC based on a nominal system model.
**for** each episode **do**
  $\xi \sim \rho_{\xi}$ randomly sample dynamics.
  $E \leftarrow \Xi(\xi)$ randomly selected environment.
  $u_{RL} \leftarrow \mu_{\theta}$ RL control input from the deterministic actor policy.
  $u = u_{RL} + u_{MBC}$ combination of the RL policy $u_{RL}$ and model-based control $u_{MBC}$.
  Initialize a random process $\mathcal{N}$ for exploration policy.
  Observe initial state.
  Perform control according to the exploration noise, RL agent action, and model-based control.
  Observe new state $s_{t+1}$ and obtain reward $r_t$.
  Store transition $(s_t, a_t, r_t, s_{t+1})$ in $R$.
  **for** each gradient step **do**
    Sample a random mini-batch of $M$ transitions $(s_i, a_i, r_i, s_{i+1})$ from $R$.
    Compute the internal memories of the LSTM actor and critic networks.
    Set $Y_i = r_i + \gamma Q_{tn}{}^{\phi_{tn}}\left(s_{i+1}, \mu_{\theta_{tn}}^{tn}(s_{i+1})\right)$.
    Update the critic parameters by minimizing the mean squared error:
    $$L(\phi) \approx \frac{1}{M}\sum_{i}\left(Q^{\phi}(s_i, a_i) - Y_i\right)^2$$
    Update the actor policy parameters using the sampled policy gradient:
    $$\nabla_{\theta} J(\mu_{\theta}) \approx \frac{1}{M}\sum_{i} \nabla_a Q^{\phi}(s, a)\big|_{s=s_i, a=\mu_{\theta}(s_i)} \nabla_{\theta}\mu_{\theta}(s)\big|_{s_i}$$
    Update the target networks:
    $\phi_{tn} \leftarrow \eta\phi + (1-\eta)\phi_{tn}$
    $\theta_{tn} \leftarrow \eta\theta + (1-\eta)\theta_{tn}$
    $u = u_{RL} + u_{MBC}$ update the hybrid control.
  **end**
**end**

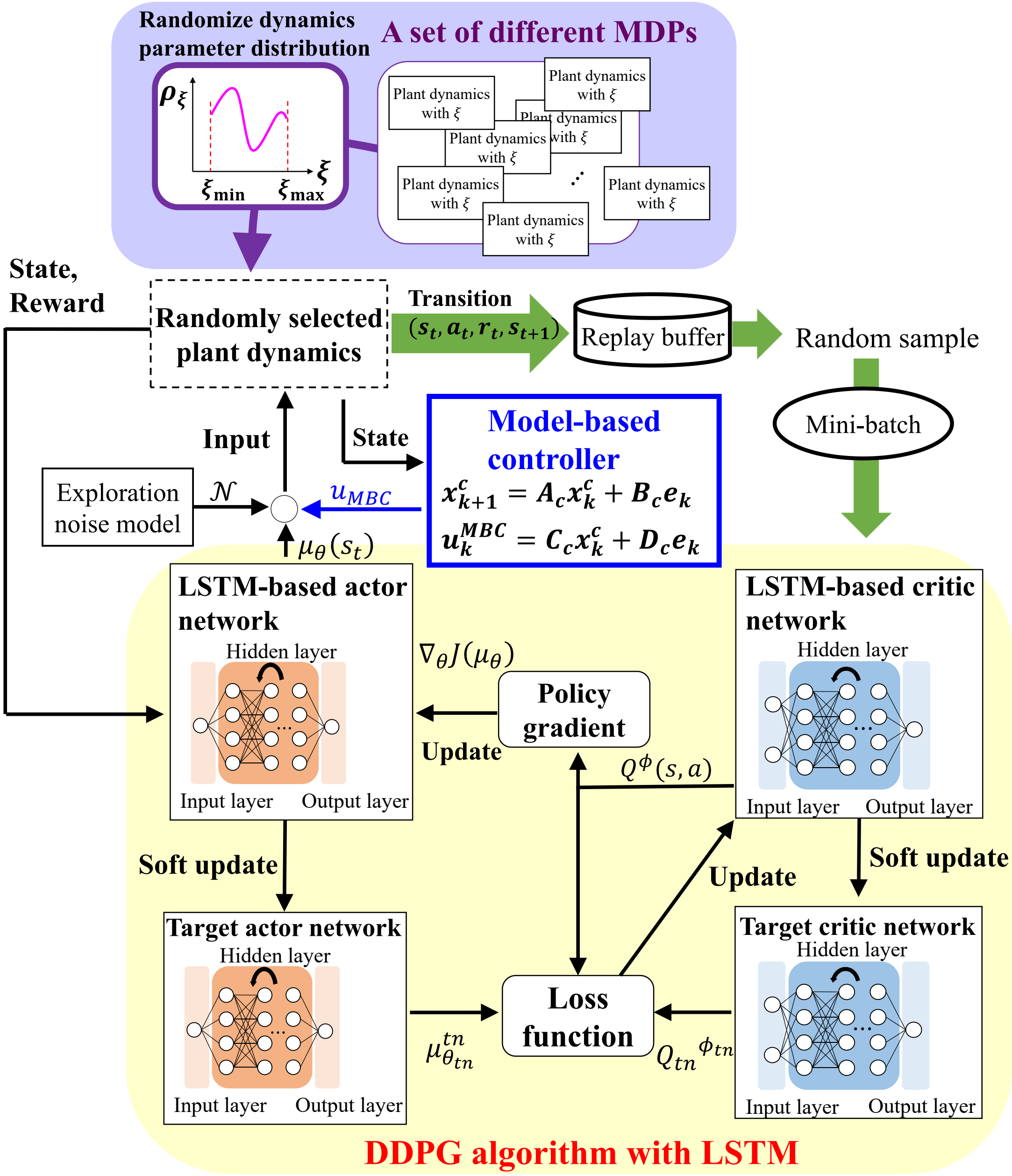

Fig. 3 DDPG algorithm with MBCA-DRT.

## 5. Numerical example

### 5.1. Active vibration control of nonlinear powertrain system

This study validates the proposed approach (i.e., MBCA-DRT) via application to active damping of a vehicle powertrain system, which suffers from nonlinearity and parametric uncertainty. Previous studies [55][56] have established a powertrain model illustrated in Fig. 4(a), where the backlash nonlinearity (i.e., dead-zone effect of a discontinuous mechanical gap) increases the vibration amplitude. The nominal values of the system parameters are listed in Table 1. As shown in Fig. 4(b), due to the presence of backlash, the dynamics are discontinuously changed between two modes, *backlash mode* and *contact mode*, making the application of traditional control more challenging. In addition, the powertrain system has parametric variations with respect to the actuator mass $M_E \in [M_E^{min},\ M_E^{max}]$ and the vehicle body mass $M_B \in [M_B^{min},\ M_B^{max}]$. Under the effects of such uncertainties, the control objective is to quickly suppress transient vibrations $X_B$ induced in the vehicle body $M_B$ by applying the control command $u$ to the actuator $M_E$. The controlled output is $y_k = X_B$, which must converge to an ideal response (i.e., reference signal ) $y_k^r$ such that $\lim_{k\to\infty} \|e_k\| = 0$, with $e_k = y_k^r - y_k = y_k^r - X_B$.

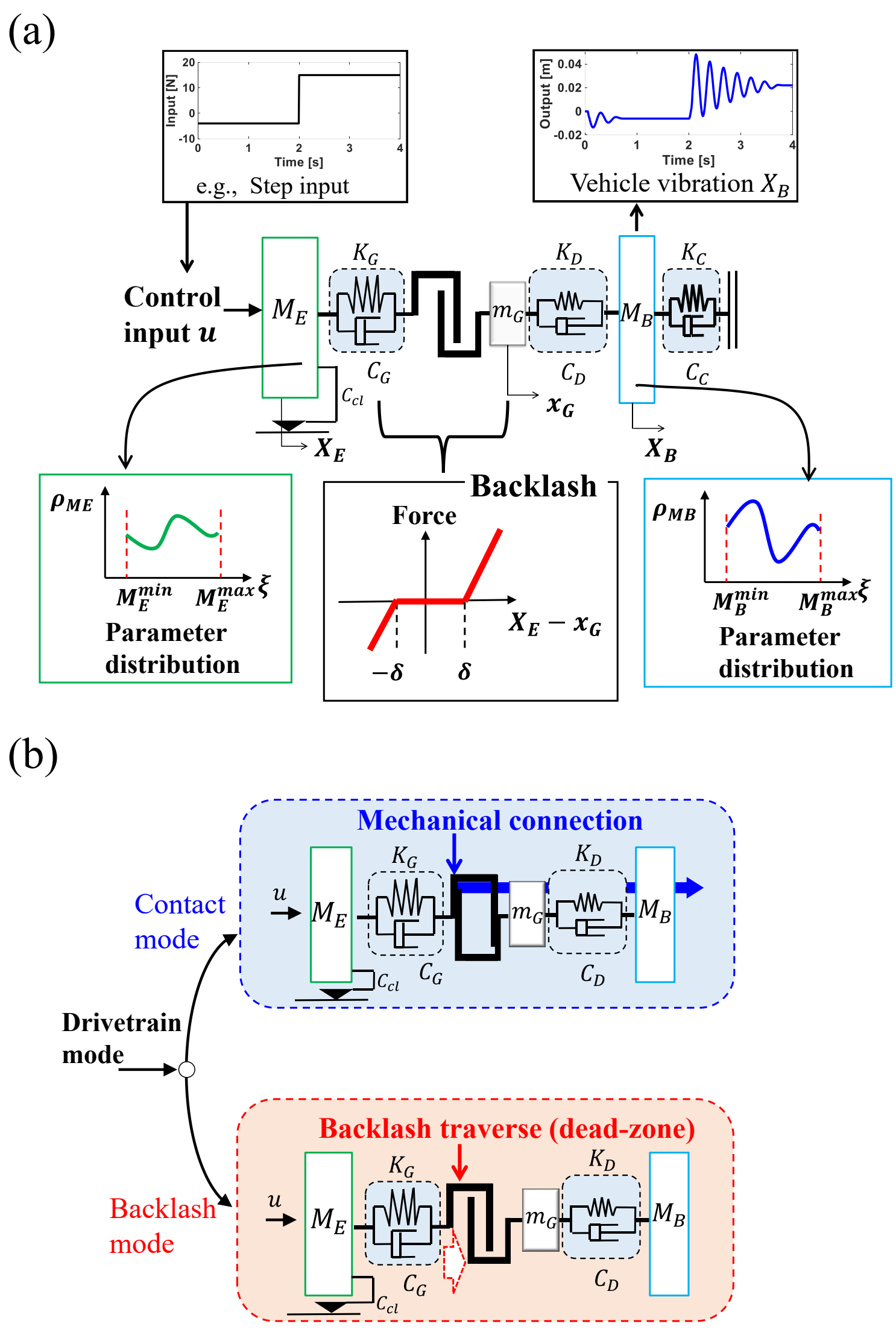

Fig. 4 Controlled system: (a) powertrain system with nonlinear backlash and parametric uncertainty and (b) two dynamics modes.

In order to design a MBC in Eqs. (7) and (8), a linearized model, which simplifies the powertrain system in Fig. 4(a) by neglecting backlash property and parameter variations, is necessary. The detailed model is available in previous studies [55][56]. Using the linearized nominal powertrain model of the previous study, this study designs an output feedback $H_2$ controller according to model-based linear control theory [57][58]. For application to powertrain control, a feedforward input is also added to the $H_2$ controller [55][56].

Table 1 Nominal parameters of the powertrain system.

| Parameter | Description | Nominal Value | Unit |
|---|---|---|---|
| $K_C$ | Spring connected with $M_B$ | 660.0 | N/m |
| $K_G$ | Spring between $M_E$ and $m_G$ | $5.3 \times 10^4$ | N/m |
| $K_D$ | Spring between $M_B$ and $m_G$ | $2.2 \times 10^4$ | N/m |
| $M_E$ | Mass of the actuator | 1.04 | kg |
| $m_G$ | Mass of the intermediate part | 0.039 | kg |
| $M_B$ | Mass of the vehicle body | 0.232 | kg |
| $C_D$ | Damper between $M_B$ and $m_G$ | 12.5 | Ns/m |
| $C_C$ | Damper connected with $M_B$ | 0.1 | Ns/m |
| $C_{cl}$ | Damper of $M_E$ | 1.5 | Ns/m |
| $C_G$ | Damper between $M_E$ and $m_G$ | 36.0 | Ns/m |
| $\delta$ | Length of the backlash | 0.005 | m |
| $y^r_{0-2s}$ | Steady value of the reference signal from 0 to 2 seconds | $-0.006$ | m |
| $y^r_{2-4s}$ | Steady value of the reference signal from 2 to 4 seconds | 0.0227 | m |

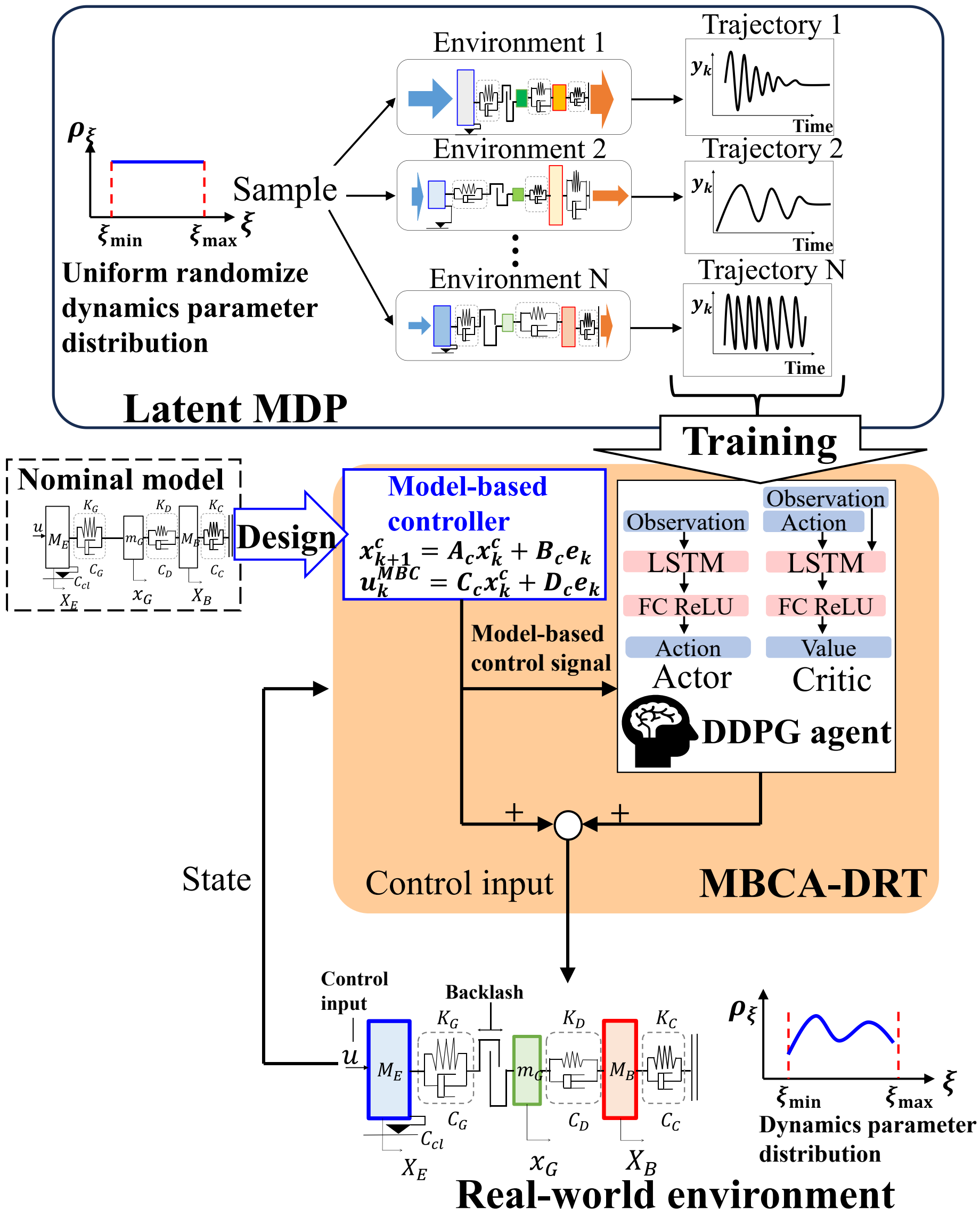

Fig. 5 Application of MBCA-DRT to active vibration control of a nonlinear powertrain system.

### 5.2. Verification settings

To examine the applicability of domain randomization, we prepare the following two scenarios. In *Scenario 1*, the actuator mass $M_E$ and the vehicle body mass $M_B$ are randomized at the beginning of each episode, whereas the amount of backlash length $\delta$ is fixed. In *Scenario 2*, which is a more challenging setup, the backlash length $\delta$ is also randomized, in addition to the two masses. Among the physical parameters of the powertrain, $M_B$, $M_E$, and $\delta$ have particularly significant effects on the system dynamics; hence their randomization is essential for robustness. This is supported by prior studies [55][59] that employed the same powertrain model. In actual vehicles, $M_B$ can change drastically with passenger load and cargo, and $\delta$ varies due to aging and manufacturing differences.

Moreover, both scenarios involve randomization of a reference signal $y_k^r$, to which the controlled output should converge.

The set-up for RL is as follows: the observed information includes the reference signal, the controlled output $y_k = X_B$, the tracking error $e_k$, its integral, and its derivative. In MBCA-DRT, the control input $u_{MBC}$ from MBC is additionally leveraged in the observation, as described in the previous chapter. Each component of the observation is properly normalized to facilitate effective learning. The reward function $r$ is defined based on the quadratic cost $c(x_k, u_k)$ in Eq. (4) as: $r = -c(x_k, u_k)$. In the simulations, the weighting matrices of the quadratic cost function in Eq. (4) were selected as $Q = 1.0$ and $R = 1.0$. In general, a larger $Q$ emphasizes tracking accuracy, while a larger $R$ emphasizes control effort saving. In this study, a simple choice of $Q = R = 1.0$ was adopted to balance both aspects.

The hyperparameters for DDPG are available in Table 2. In many prior works [60] applying DDPG, the following hyperparameters of DDPG are crucial: actor learning rate, critic learning rate, discount factor, soft target update rate (i.e., target smooth factor), batch size, and exploration noise parameters (e.g. Ornstein–Uhlenbeck noise sigma). As a rough heuristic, smaller learning rates tend to improve training stability but slow convergence. Larger batch sizes may stabilize gradient estimates but increase computational burden. A discount factor close to 1 encourages long-term reward consideration, whereas a lower value prioritizes immediate gains. The soft update coefficient is often kept small (e.g. 0.001) to slowly track target networks for stability. Increasing Ornstein–Uhlenbeck noise sigma promotes exploration in search process.

During the training process shown in Fig. 5, the above-mentioned parameters in the simulator are simultaneously randomized in each episode to allow for the RL agent to generalize for various environments. Each parameter is randomly changed at the beginning of each episode, according to the uniform distributions such as $U\left(M_B^{min}, M_B^{max}\right)$, $U\left(M_E^{min}, M_E^{max}\right)$, and $U(\delta_{min}, \delta_{max})$. Table 3 shows each variation amount. MATLAB R2024b was consistently used for both training the RL agent and verifying its performance. The specification of the PC was as follows: LAPTOP-STP755NN (ThinkPad, Lenovo Corporation), Windows 10 Pro, 64bit operating system, x64 base processor, Intel(R) Core(TM) i7-10850H CPU @ 2.70GHz 2.71 GHz.

In the numerical verifications, we compare the proposed approach (MBCA-DRT) with a case employing only a model-based linear $H_2$ controller (referred to as *Only MBC*). Note that *Only MBC* includes no compensation for backlash. Moreover, MBCA-DRT is also compared with conventional DRL-based control (referred to as *Only DR*) in which domain randomization does not receive any assistance from a MBC.

Table 2. Hyperparameters for DDPG.

| DDPG | |
|---|---|
| Hyperparameter | Value |
| Sampling time | $0.006$ s |
| Critic learning rate | $1.0 \times 10^{-4}$ |
| Actor learning rate | $1.0 \times 10^{-4}$ |
| Discount factor $\gamma$ | $0.99$ |
| Exploration noise $\mathcal{N}$ | Ornstein-Uhlenbeck Action Noise |
| Number of the neurons in the hidden layers | $64$ |
| Size of the mini-batch $M$ | $128$ |
| Size of the replay buffer | $1.0 \times 10^{5}$ |
| Target smooth factor $\eta$ | $1.0 \times 10^{-3}$ |
| Horizon $T$ | $667$ |
| Network architecture | LSTM + Fully connected + ReLu |
| Number of episodes | $300$ |

Table 3. Randomized dynamics parameters.

| Parameter | Range |
|---|---|
| Vehicle body mass: $M_B$ | Nominal value $\pm 50\%$: $\left[M_B^{min}, M_B^{max}\right] = [0.1160, 0.3480]$ |
| Actuator mass: $M_E$ | Nominal value $\pm 50\%$: $\left[M_E^{min}, M_E^{max}\right] = [0.5200, 1.5600]$ |
| Length of backlash: $\delta$ (randomized only in *Scenario 2*) | Nominal value $\pm 50\%$: $[\delta_{min}, \delta_{max}] = [0.0025, 0.0075]$ |
| Steady value of the reference signal from 0 to 2 seconds: $y_{0-2s}^r$ | $\left[y_{0-2s}^{r\,min}, y_{0-2s}^{r\,max}\right] = [-0.01515, -0.00151515]$ m |
| Steady value of the reference signal from 2 to 4 seconds: $y_{2-4s}^r$ | $\left[y_{2-4s}^{r\,min}, y_{2-4s}^{r\,max}\right] = [0.0151515, 0.030303]$ m |

### 5.3. Results and discussion in *Scenario 1*

Figure 6 shows reward curves in *Scenario 1* and *Scenario 2*. The red and blue lines describe the episode rewards obtained by the proposed method (i.e., MBCA-DRT) and *Only DR*, respectively. As shown in the figure, the proposed method exhibits three clear advantages: (1) the exploration process is stabilized within the first few episodes; (2) convergence is achieved with significantly fewer episodes; and (3) the final reward level is higher than that of domain randomization alone. These results confirm that the proposed method enables more efficient and effective training. By ensuring a baseline control performance from the beginning, MBC prevents the RL agent from engaging in unnecessary exploration.

As a result, the agent can focus on refining the control strategy around an already stable baseline, which substantially accelerates the learning process.
It should also be noted that the reward curve of the proposed method exhibits a temporary drop around 200 episodes in Fig. 6(b). Such sudden declines may be due to factors such as exploration noise, distributional shifts in the replay buffer, or temporary critic instability. Nevertheless, thanks to the baseline performance ensured by MBC, the agent quickly recovers from this transient collapse and continues to converge to a stable and superior performance compared to domain randomization alone.

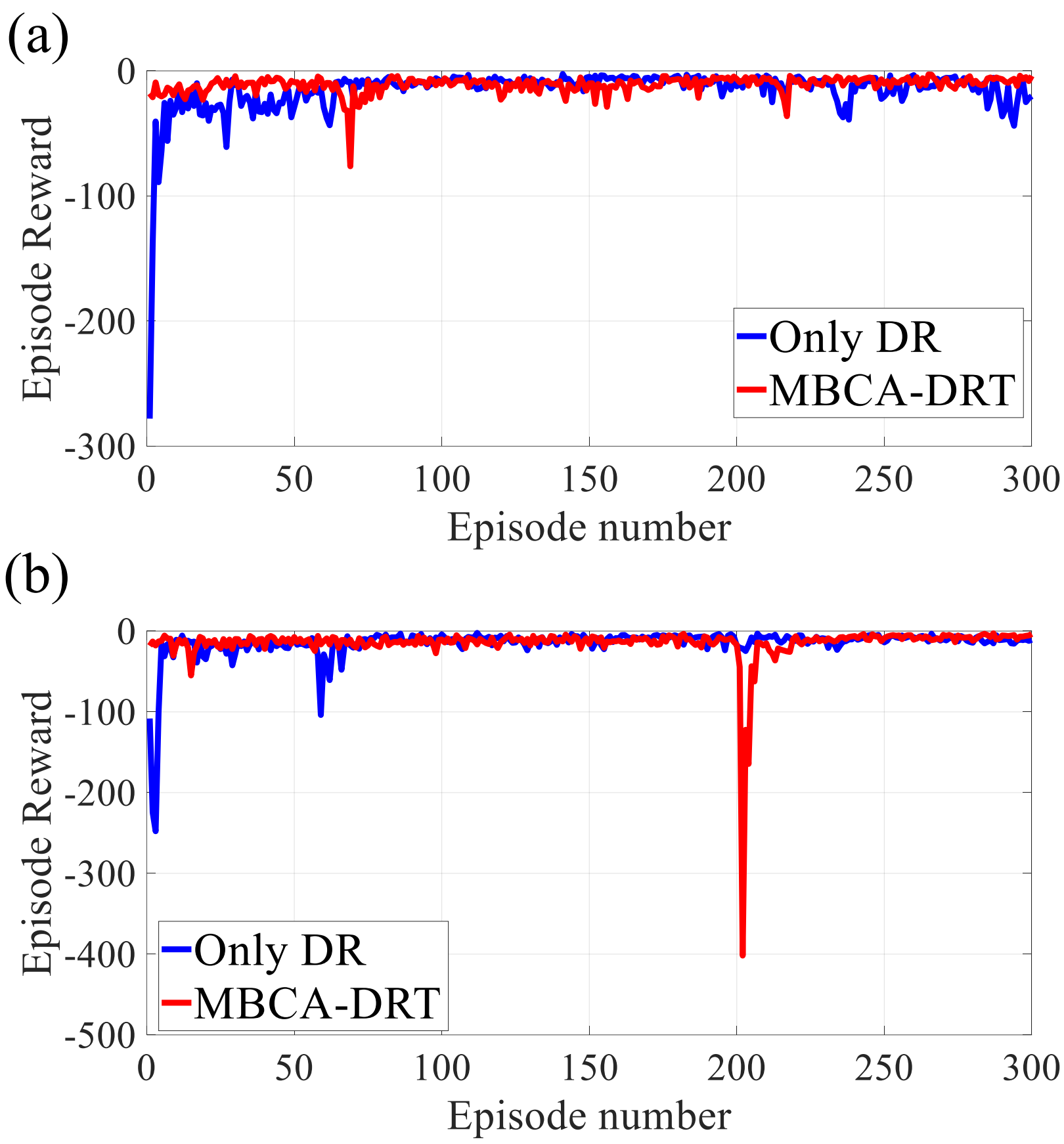

Fig. 6 Reward curves in (a) *Scenario 1* and (b) *Scenario 2*.

First, we consider *Scenario 1*. Figure 7 shows the control result for a nominal powertrain system without any parametric variations and backlash nonlinearity. This work quantitatively evaluates the control performance of each control based on 2-norm of the tracking error $e_k = y_k^r - y_k$, corresponding to the first term in Eq. (4). These results are summarized in Table 4. Figure 7 indicates that all the control systems (proposed approach, *Only DR*, and *Only MBC*) successfully damp the vehicle body vibrations, making the controlled output quickly converge to the target response. According to Table 4, the minimum norm is achieved by *Only MBC*. It is because there are no sim-to-real gaps between the nominal model used for controller design and the real system, which means that the MBC should be the best choice.

Figures 8-11 show the control results for powertrain systems with nonlinearity and parameter uncertainty. Though we examined more cases by changing patterns of variations of the randomized parameters for *Scenario 1*, only the representative results are given here. Those 2-norm values are available in Tables 5-8. The remarkable point is that the proposed approach denoted in the red line consistently maintains the high control performance, proving its excellent robustness and generalization capabilities to real testing environments. Even with the parametric variations in Figs. 8-11, the transient vibrations are attenuated. In addition, the overshoots owing to the effect of backlash are reduced immediately after 2.0 s compared to the cyan and black lines. Consequently, the proposed approach can comprehensively deal with parametric uncertainty and dynamical nonlinearity.

In contrast, the residual vibration remains in the responses by *Only DR*, even though instability of the controlled output is avoided in all cases. The cause of such poor performance lies in the training process where the learning convergence is immature. The immature training procedure may require more data (i.e., episodes) or larger network structures, whereas the proposed approach indicates that the current small setup is sufficient to achieve the control objective. This comparison clearly demonstrates that the introduction of MBC can accelerate the learning process. Despite the system variations, responses were stabilized with *Only DR* due to the conservativeness of the policy. In other words, the RL agent is so overfitting the policy to avoid worst case (i.e., unstable response) that it also loses its aggressiveness to suppress transient vibrations.

Focusing on the responses by *Only MBC*, we can see that the variation in control performance is more drastic, as implied from comparison of Figs. 8 and 11. Specifically, Table 8 presents the good performance achieved by *Only MBC*. In contrast, the response is on the verge of destabilization in Fig. 8. Since MBC relies on the approximate linear model, robustness cannot be ensured if the sim-to-real gap is not considered explicitly. In all cases, the overshoots largely remain, which is due to the lack of compensation for backlash nonlinearity.

In summary, the comparative verification between the proposed approach, *Only MBC*, and *Only DR* shows the effectiveness of the combination of domain randomization and MBC. This combination prevents the controller from being too conservative while improving the robustness.

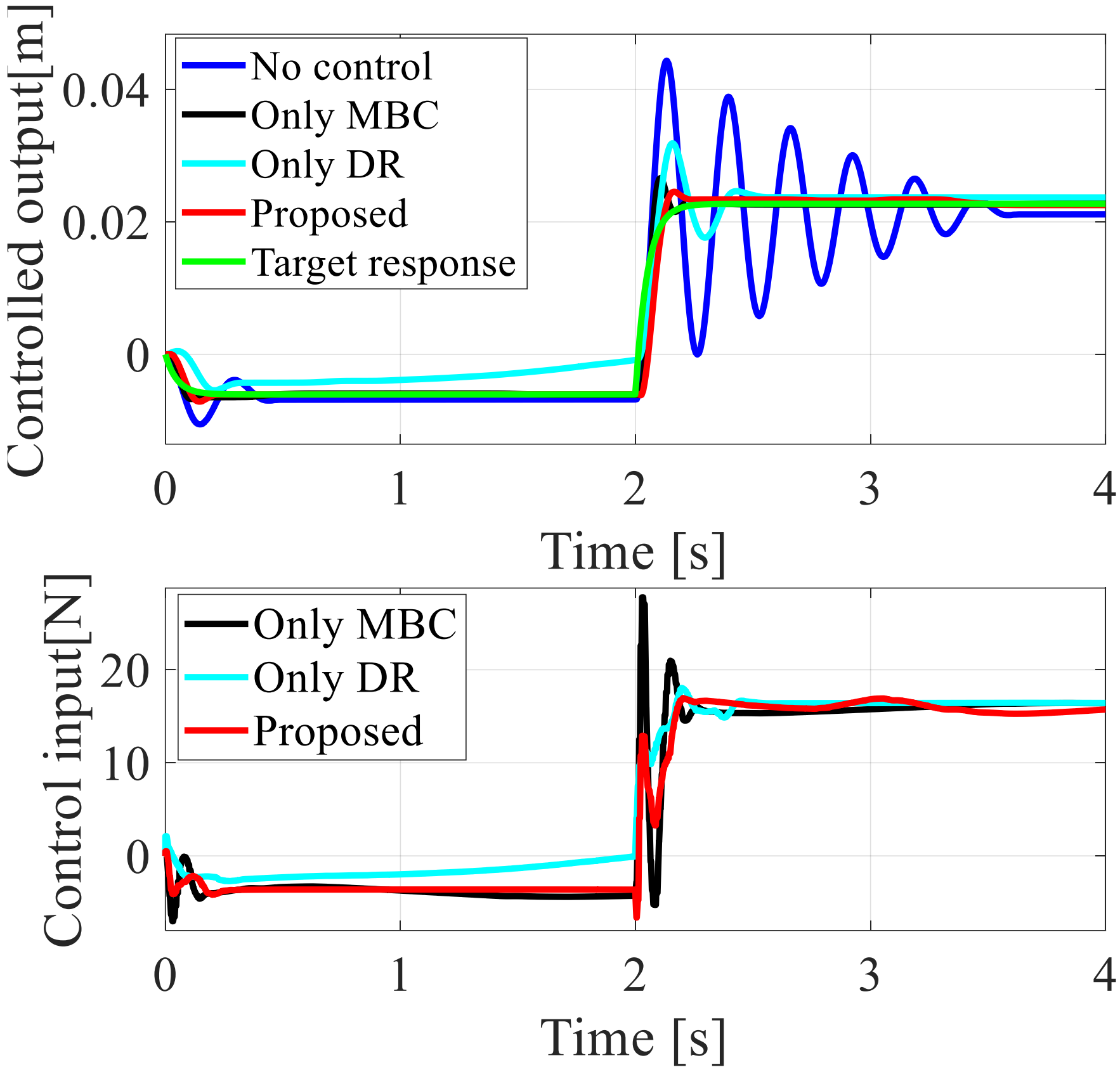

Fig. 7 Time responses of the vehicle body vibration (upper graph) and the control input (lower graph) in *Scenario 1*, where all the parameters take nominal values and backlash is absent in the controlled powertrain.

Table 4 2-norm computed for the control results in Fig. 7.

| | Proposed method (MBCA-DRT) | Only domain randomization | Only MBC | Open-loop (i.e., no control) |
|---|---|---|---|---|
| 2-norm | 0.7523 | 1.2311 | 0.5890 | 2.6940 |

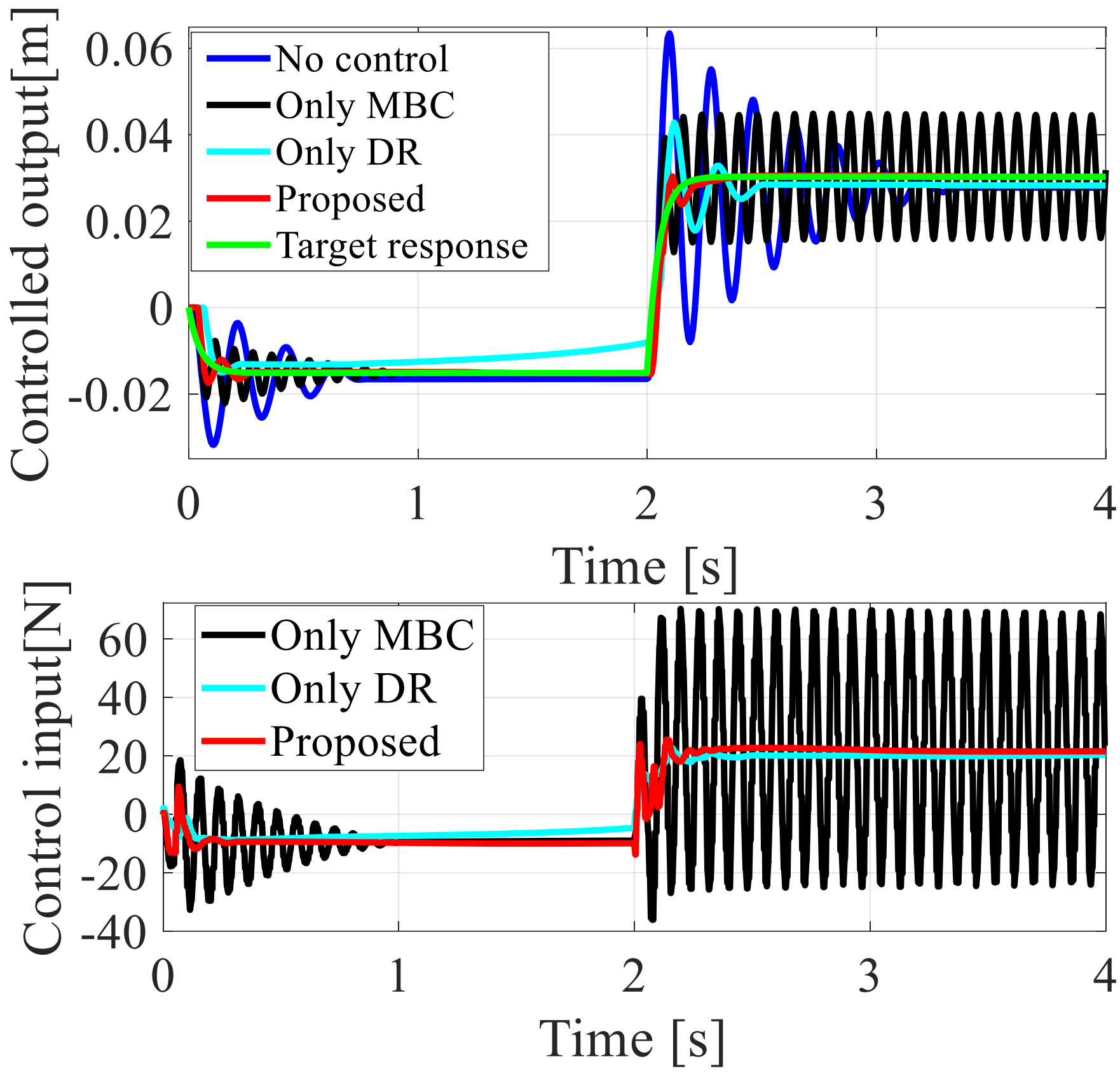

Fig. 8 Time responses of the vehicle body vibration and the control input in *Scenario 1* with $M_B = M_B^{min}$, $M_E = M_E^{min}$, $y_{0-2s}^r = y_{0-2s}^{r\,min}$, $y_{2-4s}^r = y_{2-4s}^{r\,max}$, and the fixed backlash length $\delta$.

Table 5 2-norm computed for the control results in Fig. 8.

| | Proposed method (MBCA-DRT) | Only domain randomization | Only MBC | Open-loop (i.e., no control) |
|---|---|---|---|---|
| 2-norm | 0.7916 | 1.7397 | 3.3337 | 3.9985 |

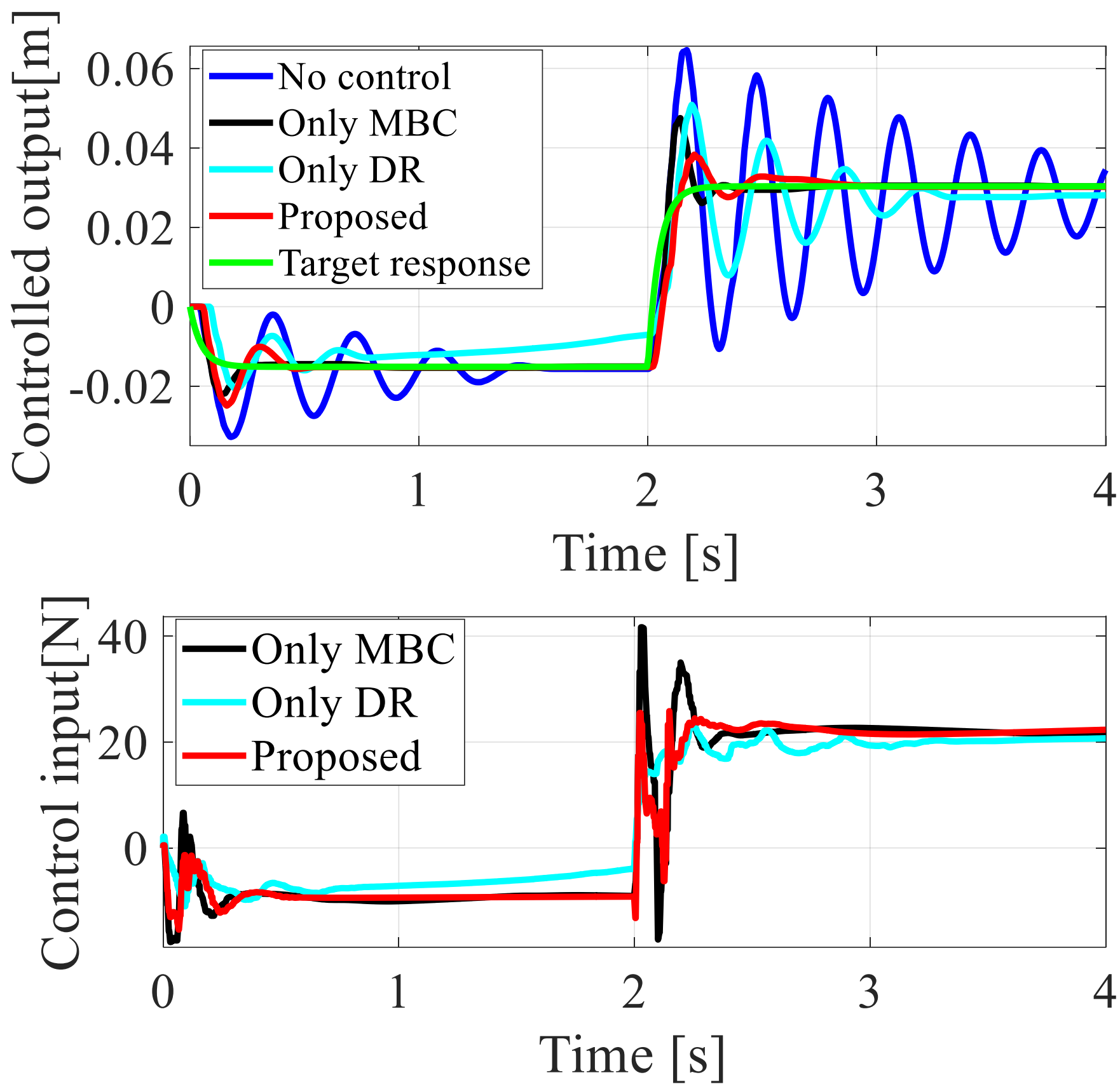

Fig. 9 Time responses of the vehicle body vibration and the control input in *Scenario 1* with $M_B = M_B^{max}$, $M_E = M_E^{max}$, $y_{0-2s}^r = y_{0-2s}^{r\,min}$, $y_{2-4s}^r = y_{2-4s}^{r\,max}$, and the fixed backlash length $\delta$.

Table 6 2-norm computed for the control results in Fig. 9.

| | Proposed method (MBCA-DRT) | Only domain randomization | Only MBC | Open-loop (i.e., no control) |
|---|---|---|---|---|
| 2-norm | 1.5540 | 3.0564 | 1.5609 | 5.8579 |

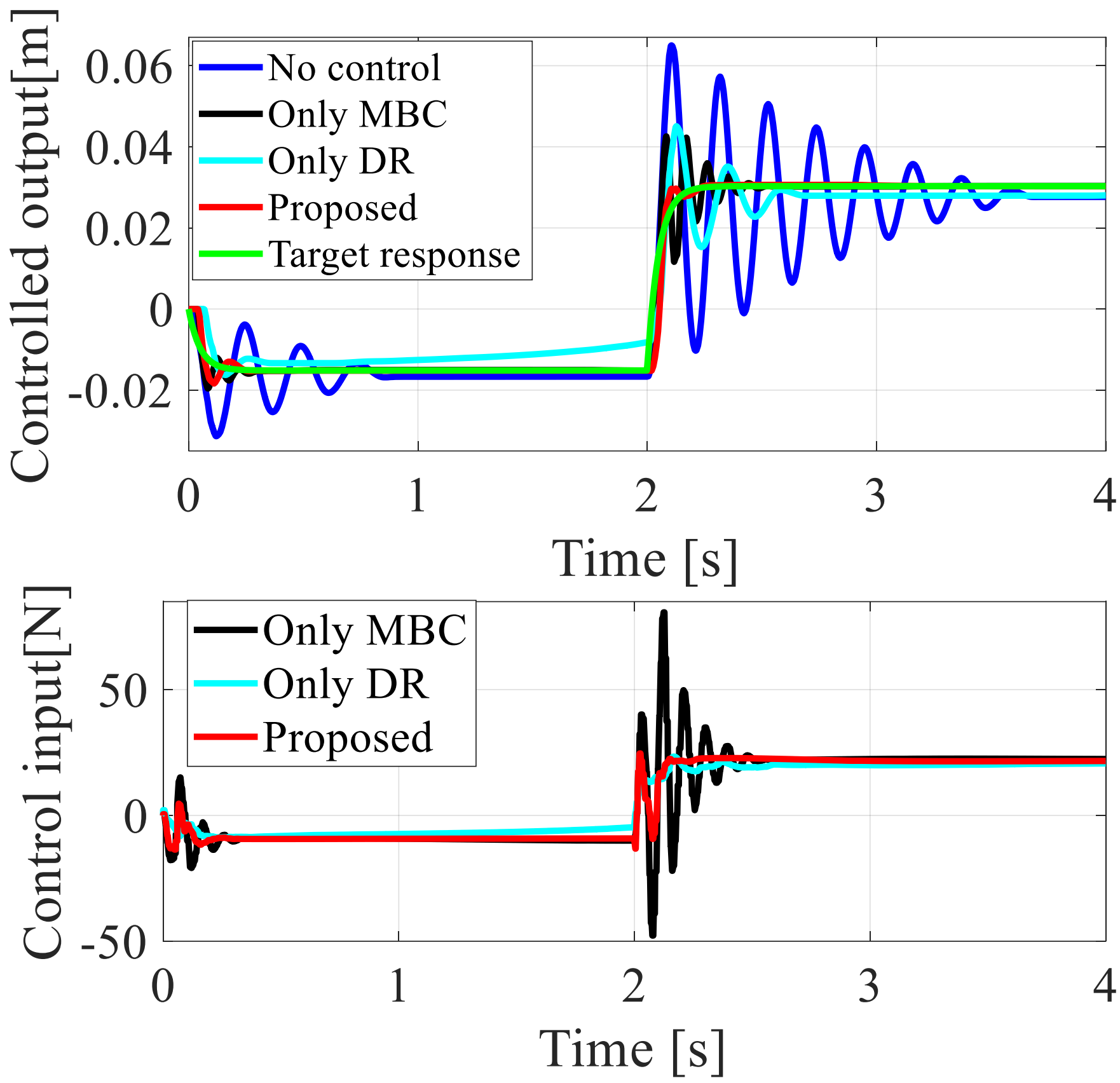

Fig. 10 Time responses of the vehicle body vibration and the control input in *Scenario 1* with $M_B = M_B^{max}$, $M_E = M_E^{min}$, $y_{0-2s}^r = y_{0-2s}^{r\,min}$, $y_{2-4s}^r = y_{2-4s}^{r\,max}$, and the fixed backlash length $\delta$.

Table 7 2-norm computed for the control results in Fig. 10.

| | Proposed method (MBCA-DRT) | Only domain randomization | Only MBC | Open-loop (i.e., no control) |
|---|---|---|---|---|
| 2-norm | 0.9505 | 1.9876 | 1.3512 | 4.5873 |

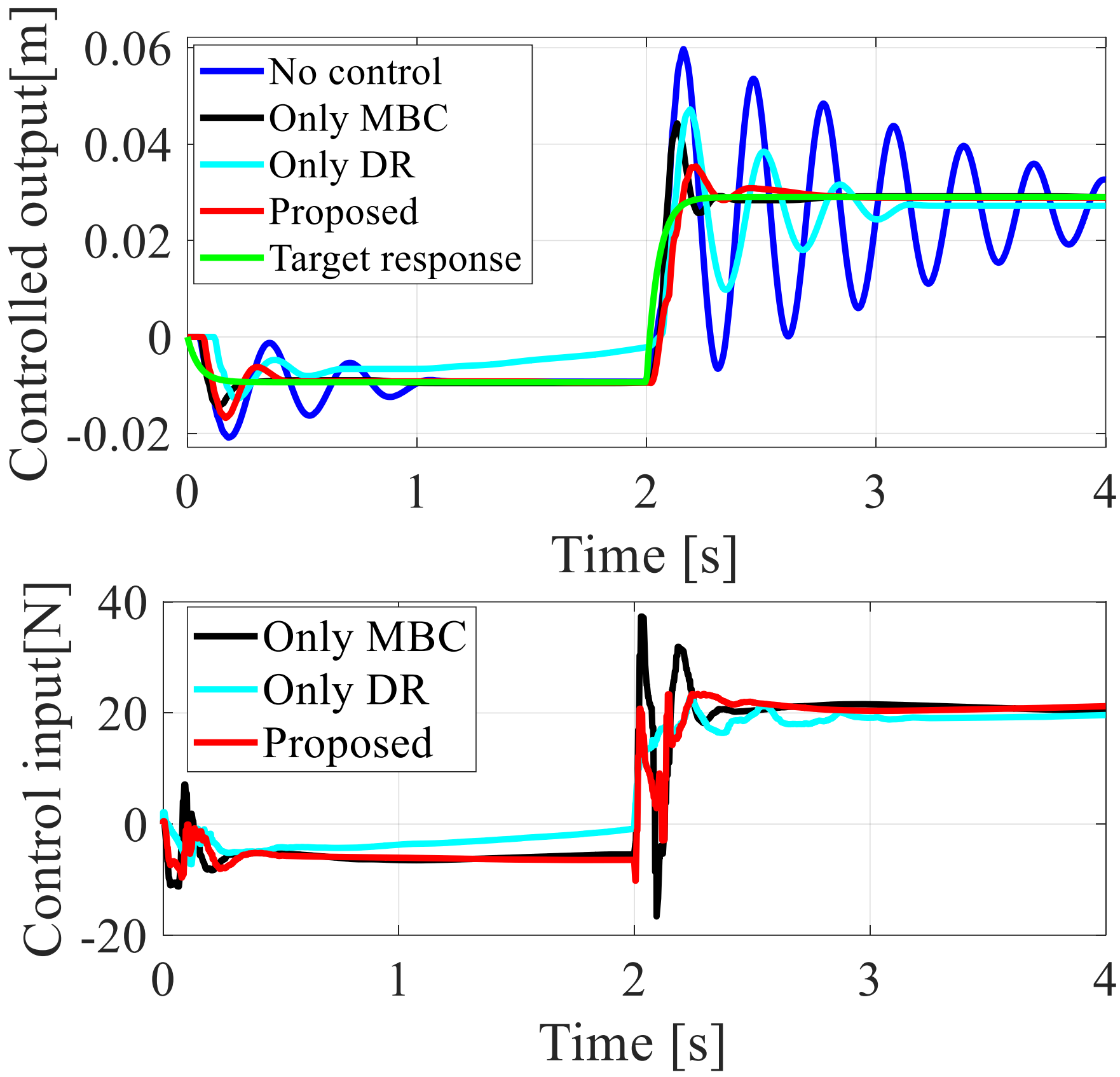

Fig. 11 Time responses of the vehicle body vibration and the control input in *Scenario 1* with $M_B = 0.2998$, $M_E = 1.5179$, $y^r_{0-2s} = -0.00940022$, $y^r_{2-4s} = 0.0290263$, and the fixed backlash length $\delta$.

Table 8 2-norm computed for the control results in Fig. 11.

| | Proposed method (MBCA-DRT) | Only domain randomization | Only MBC | Open-loop (i.e., no control) |
|---|---|---|---|---|
| 2-norm | 1.3352 | 2.5931 | 1.3173 | 4.8748 |

We also analyze the necessity of LSTM in the proposed approach. Figure 12 shows the control result obtained by the combination of MBC and domain randomization employing multi-layer perceptrons (MLPs), not LSTM. The MLP and LSTM have the same hyperparameters. The robustness has decreased, resulting in the transient vibration. The controlled powertrain system has the same conditions as those of Fig. 8. Therefore, the comparison clarifies the importance of incorporating memory-augmented networks such as LSTM for the actor and critic to allow the agent to infer the dynamics of randomized environments.

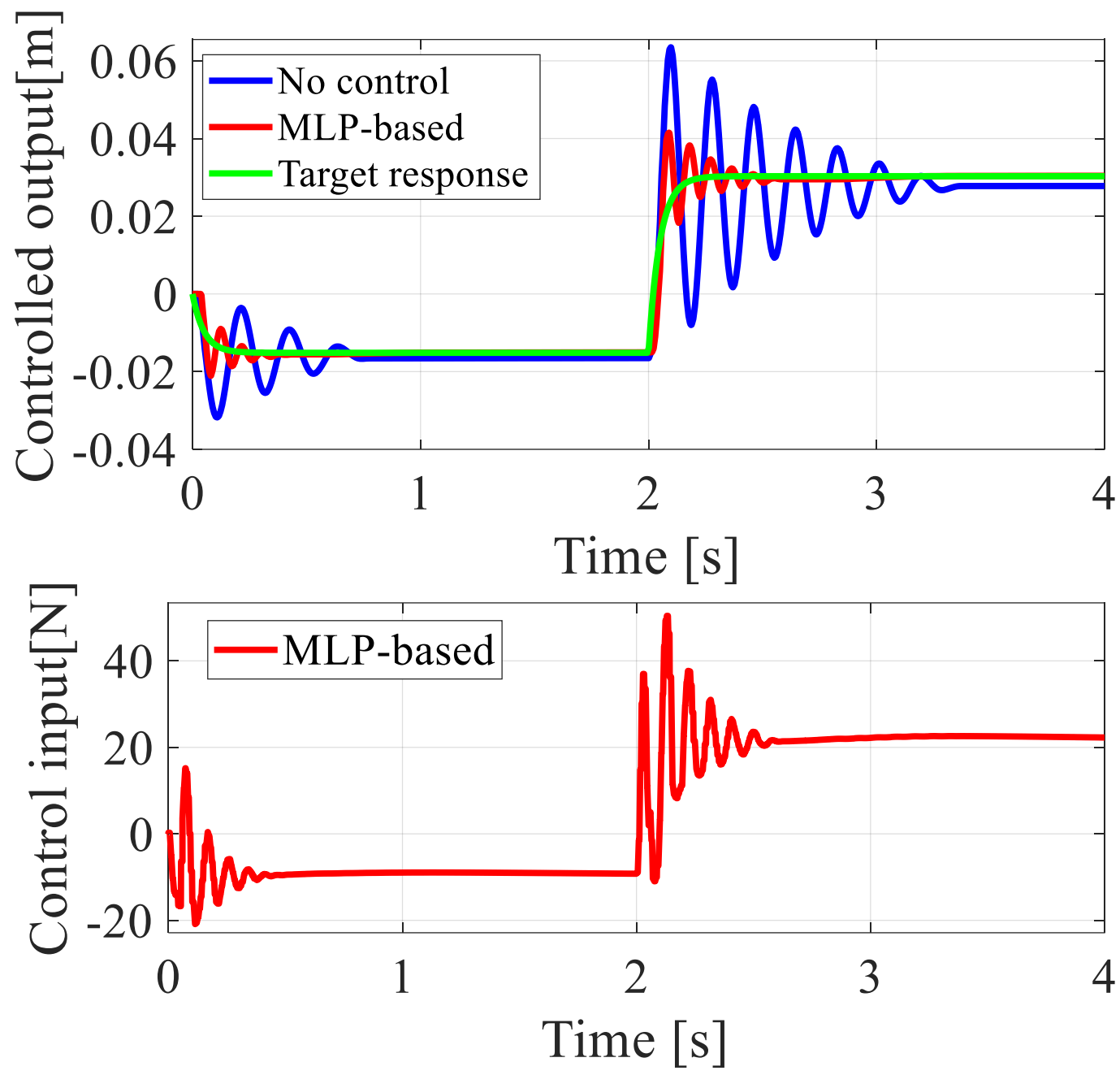

Fig. 12 Control result by multi-layer perceptron (MLP)-based actor and critic in *Scenario 1* with $M_B = M_B^{min}$, $M_E = M_E^{min}$, $y_{0-2s}^r = y_{0-2s}^{r\,min}$, $y_{2-4s}^r = y_{2-4s}^{r\,max}$, and the fixed backlash length $\delta$.

### 5.4. Results and discussion in *Scenario 2*

In *Scenario 2*, the size of backlash $\delta$ is also randomized during training. This is a more challenging setup because of the increasing number of randomized parameters.

The control result for the nominal powertrain system is illustrated in Fig. 13. The quantitative comparison is available in Table 9. Although the proposed approach completely suppresses the overshoot (i.e., the effect of backlash), the best performance is provided by *Only MBC*. As discussed in the previous section, this is a reasonable result since the controlled system has no uncertainties.

The strong robustness of the proposed approach is demonstrated through Figs. 14-16 and Tables 10-12, which indicate the test results of robustness to perturbed powertrain systems. Despite the variations in backlash length, in addition to the changed two masses, the red line exhibits the smallest overshoot as well as the reduced transient vibrations in all cases. Compared to the cyan line, we can obtain the following conclusion: *concurrently using MBC with domain randomization can ensure robustness while preventing the policy from being too conservative*. For example, the powertrain has the maximum size of backlash $\delta_{max}$ in Fig. 14, where *Only DR* fails to suppress the overshoot due to the backlash. This is because of the too conservative policy. Figure 17 shows the enlarged view of Fig. 14 immediately after 2.0 s. As indicated by the magenta arrow, the cyan waveform (*Only DR*) lacks a sufficient downward force in the negative direction, which is crucial for suppressing overshoot. In other words, the control exhibits a lack of aggressiveness because of its conservativeness. In contrast, the red line shows the opposite tendency, where the large negative force is instantaneously applied, effectively damping the overshoot.

Compared to the black line, the robustness of the proposed approach is also clearly identified. *Only MBC* shows the good performances in Figs. 15 and 16, whereas the vibration is hardly attenuated in Fig. 14. Analyzing in detail Fig. 17, excessive control input and its phase error occur in *Only MBC*. This deterioration is due to the parameter gaps of $M_E$, $M_B$, and $\delta$ to their nominal values. On the other hand, the proposed approach is consistently robust against various dynamics in Figs. 14-16. This tendency indicates the high generalization ability to real testing environments.

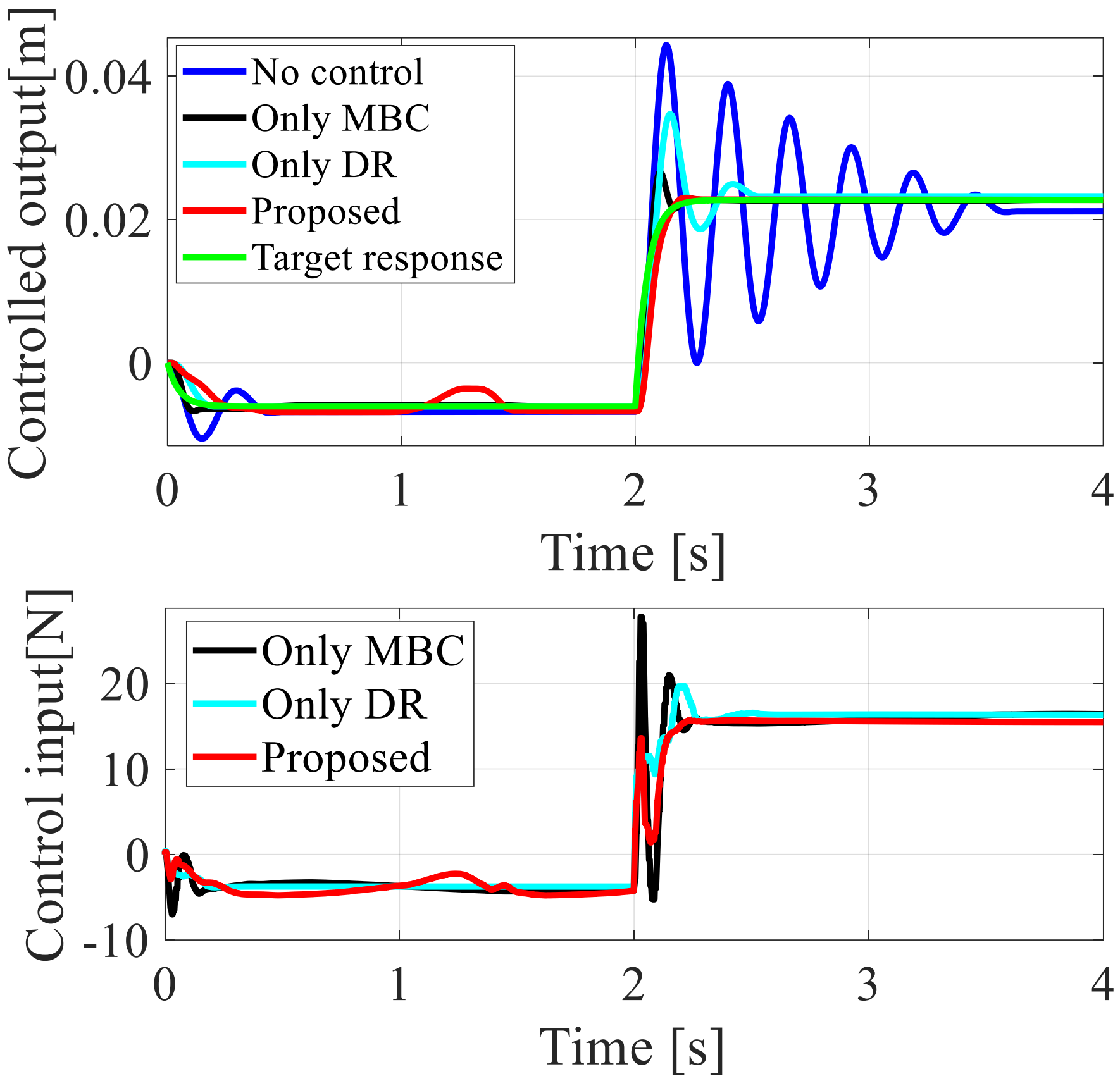

Fig. 13 Time responses of the vehicle body vibration (upper graph) and the control input (lower graph) in *Scenario 2*, where all the parameters take nominal values and backlash is absent in the controlled powertrain.

Table 9 2-norm computed for the control results in Fig. 13.

| | Proposed method (MBCA-DRT) | Only domain randomization | Only MBC | Open-loop (i.e., no control) |
|---|---|---|---|---|
| 2-norm | 0.7179 | 0.9867 | 0.5890 | 2.6940 |

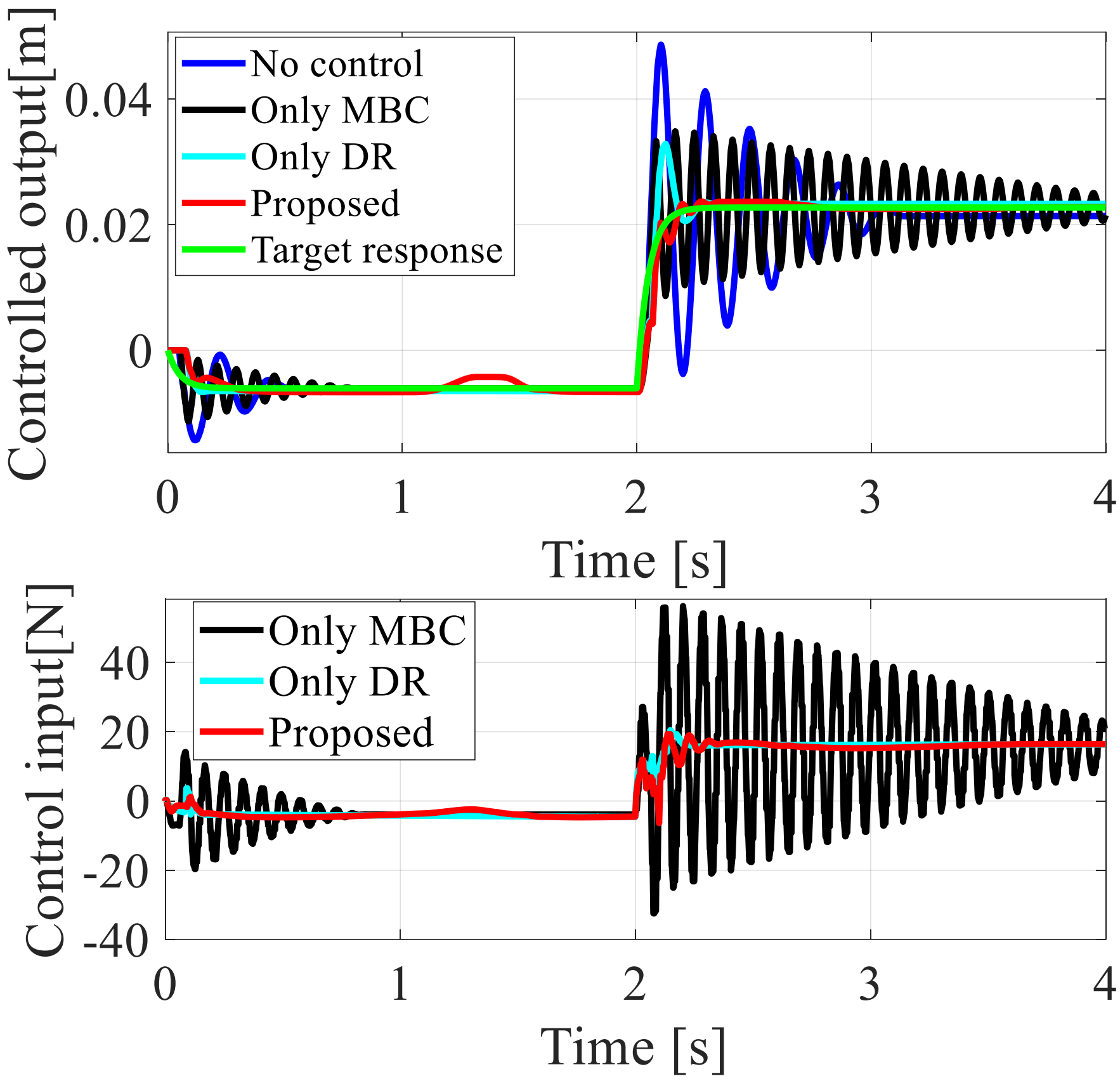

Fig. 14 Time responses of the vehicle body vibration and the control input in *Scenario 2* with $M_B = M_B^{min}$, $M_E = M_E^{min}$, $y_{0-2s}^r = -0.006$, $y_{2-4s}^r = 0.0227$, and $\delta = \delta_{max}$.

Table 10 2-norm computed for the control results in Fig. 14.

| | Proposed method (MBCA-DRT) | Only domain randomization | Only MBC | Open-loop (i.e., no control) |
|---|---|---|---|---|
| 2-norm | 0.6694 | 0.8379 | 1.8922 | 2.6610 |

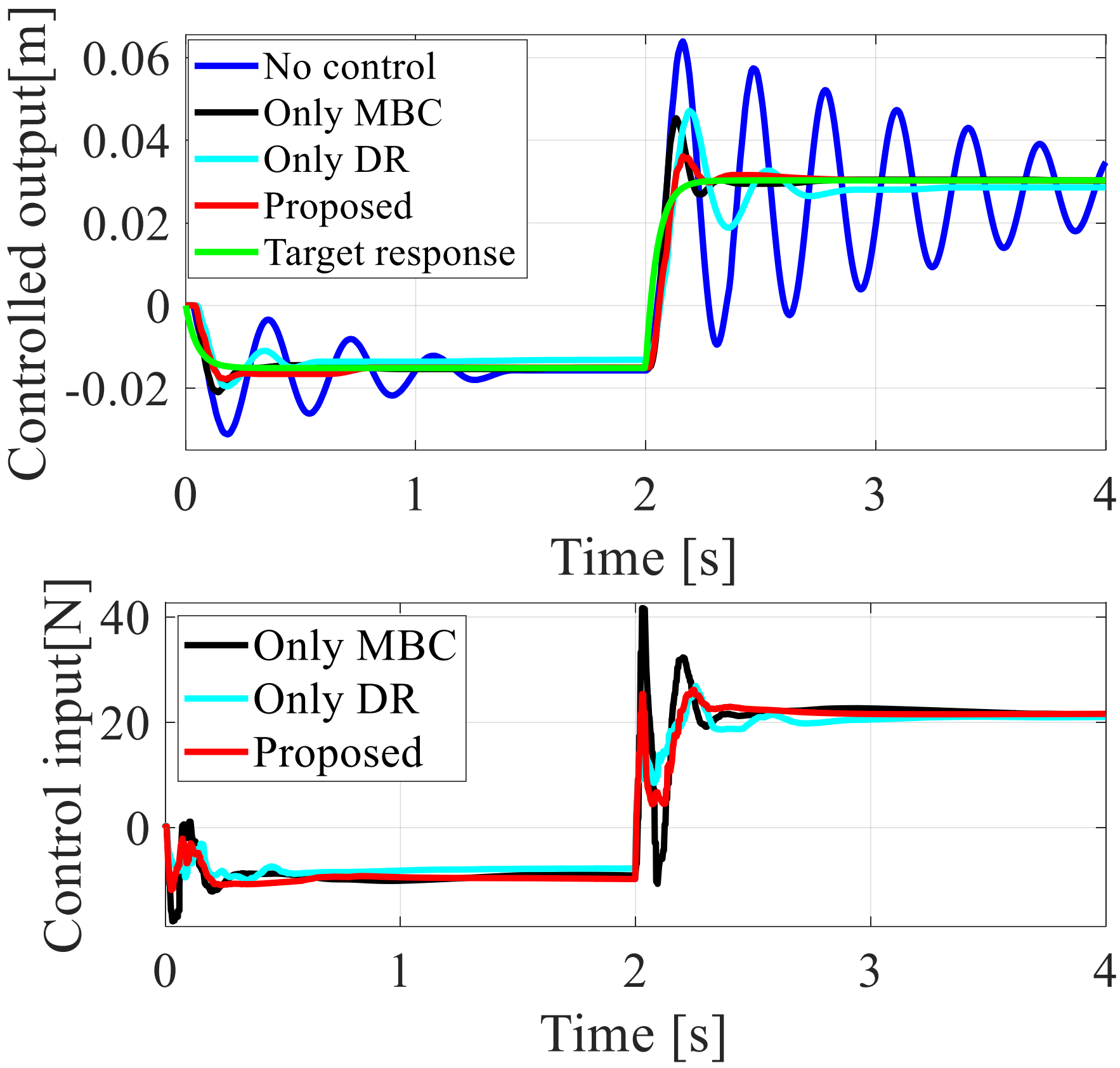

Fig. 15 Time responses of the vehicle body vibration and the control input in *Scenario 2* with $M_B = M_B^{max}$, $M_E = M_E^{max}$, $y_{0-2s}^r = y_{0-2s}^{r\,min}$, $y_{2-4s}^r = y_{2-4s}^{r\,max}$, and $\delta = \delta_{min}$.

Table 11 2-norm computed for the control results in Fig. 15.

| | Proposed method (MBCA-DRT) | Only domain randomization | Only MBC | Open-loop (i.e., no control) |
|---|---|---|---|---|
| 2-norm | 1.1789 | 1.9748 | 1.4417 | 5.6867 |

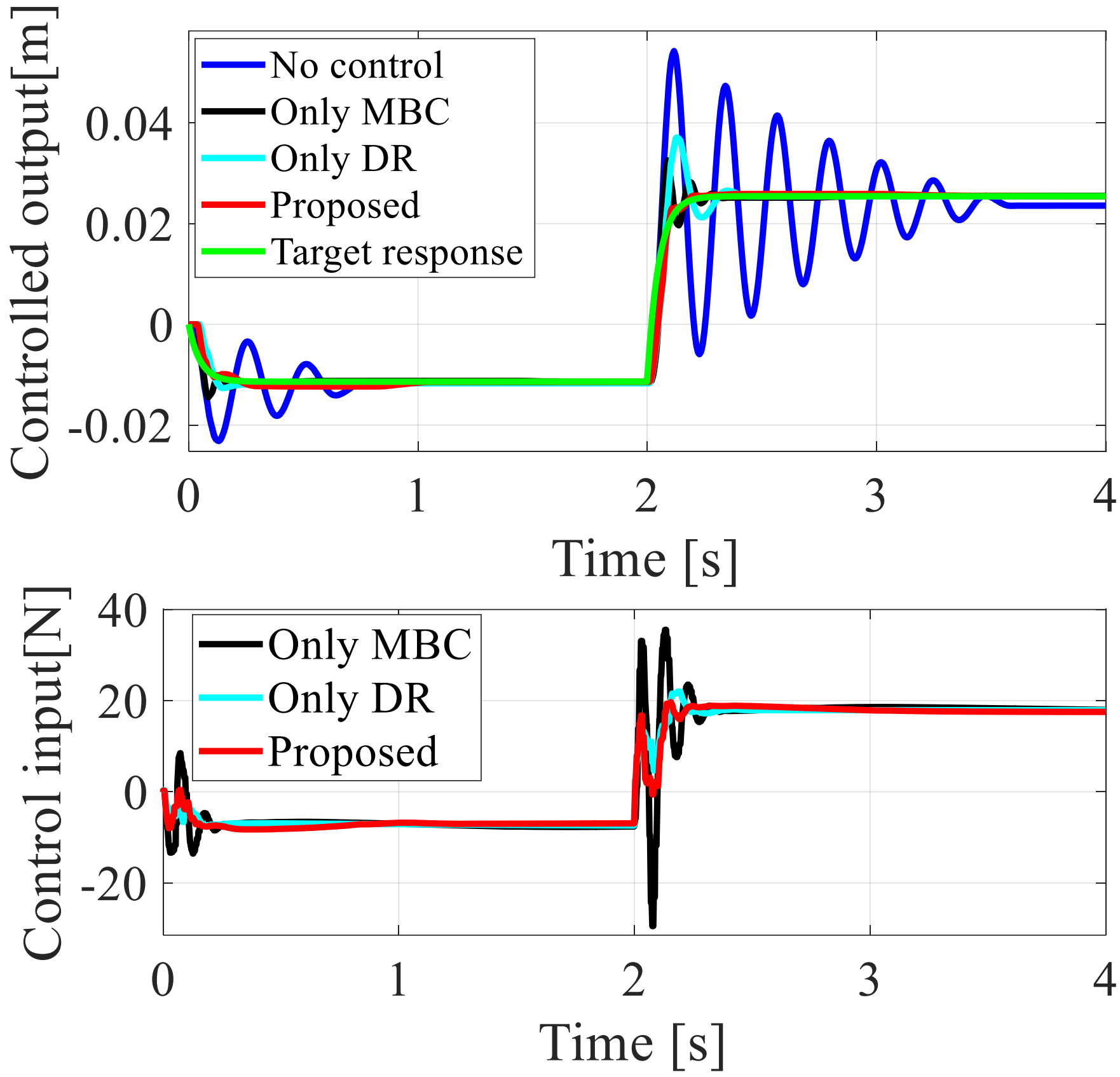

Fig. 16 Time responses of the vehicle body vibration and the control input in *Scenario 2* with $M_B = 0.26798$, $M_E = 0.68912$, $y^r_{0-2s} = -7.5158$, $y^r_{2-4s} = 16.7970$, and $\delta = 0.00309498$.

Table 12 2-norm computed for the control results in Fig. 16.

| | Proposed method (MBCA-DRT) | Only domain randomization | Only MBC | Open-loop (i.e., no control) |
|---|---|---|---|---|
| 2-norm | 0.6682 | 1.0651 | 0.8494 | 3.5979 |

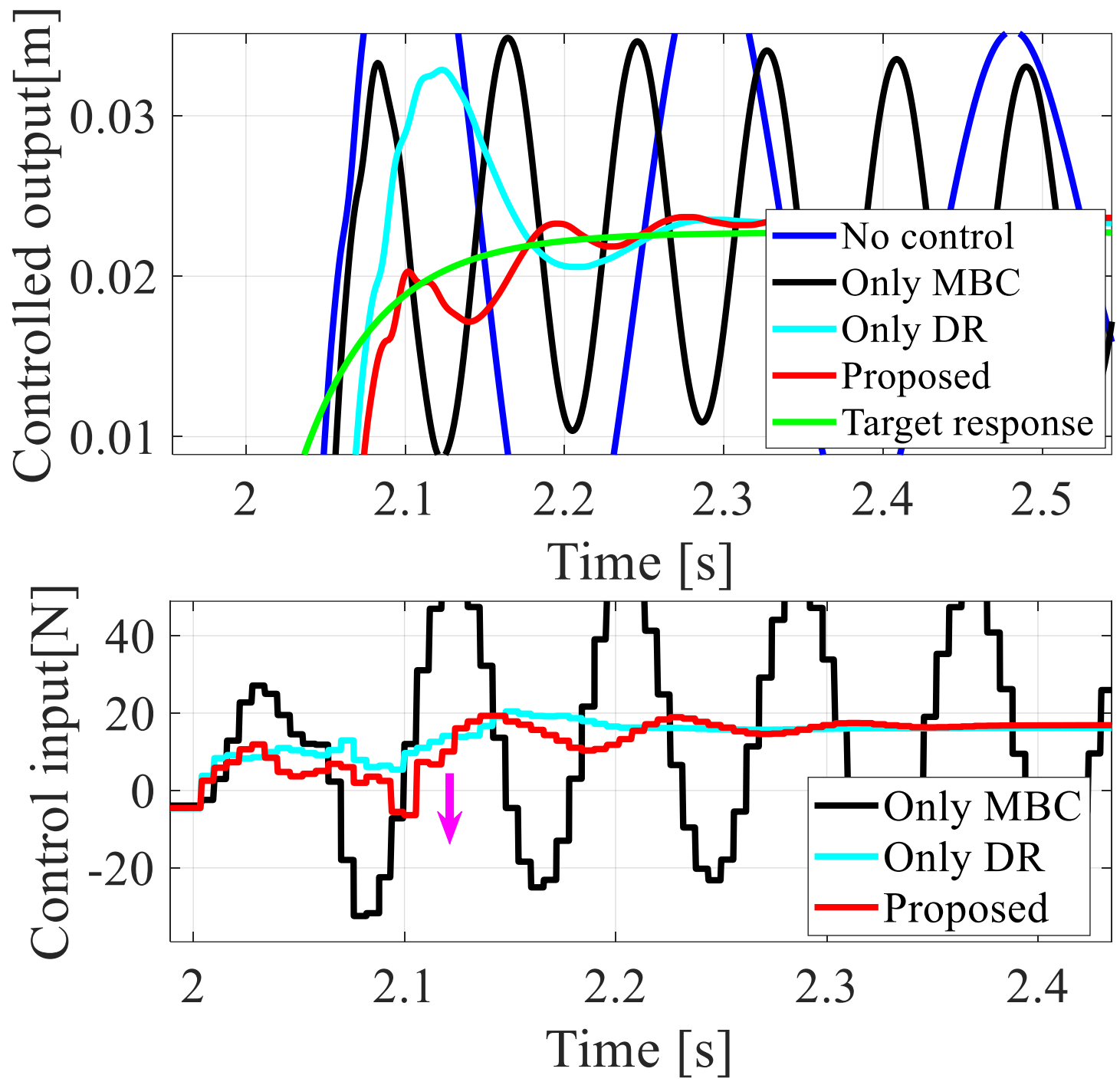

Fig. 17 Enlarged view of the response in Fig. 14.

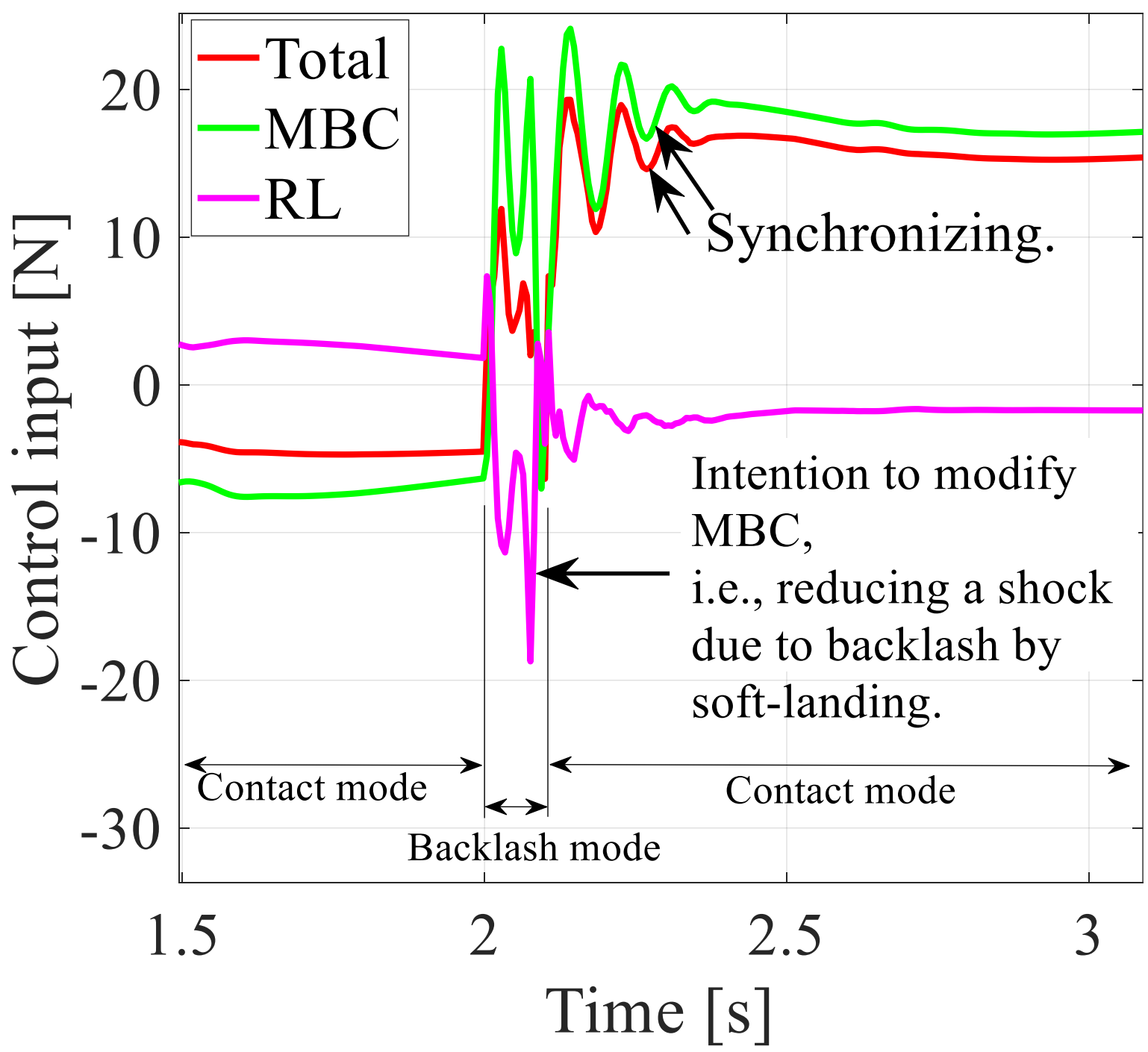

Fig. 18 Breakdown of the proposed control input in Fig. 14.

We can conduct a detailed analysis of the control input computed by the proposed approach in Fig. 18, which provides an enlarged view of the breakdown of the total control input composed of MBC and RL, focusing on the key time zone. As discussed below, Fig. 18 supports the presence of our proposed synergy effect of MBC and RL.

The first remarkable point is that the control behavior of MBC and the total control (i.e., MBC+RL) are generally synchronized, as seen from the matching of their input phases. This means that the base of the control action is formed by MBC, which is mainly responsible for constructing the total control input. As indicated by the difference in the input magnitude between the red and green lines, however, *Only MBC* results in improper control action due to the system uncertainties, e.g., backlash nonlinearity. Immediately after 2.0 s, the green line shows that MBC computes an excessive amount of control input due to the dead zone effect of backlash, which will produce shock by unnecessarily accelerating the actuator. Such a shock effect causes a large overshoot and remaining vibrations.

As shown in the magenta line, the more critical highlight is the control behavior by RL to properly modify the control input of MBC. As mentioned earlier, MBC generates an excessively large positive control input immediately after 2.0 s, which will lead to shock due to the effect of backlash, specifically during the transition from the backlash mode to the contact mode. In contrast, RL computes the negative control input in the opposite direction to counteract the excessive control input of MBC. This is evident from the phase opposition between the magenta waveform and the green waveform. The modification by RL appropriately reduces the magnitude of the total control input, indicated by the red waveform, compared to the green waveform. This contributes to the improvement in control performance.

The corresponding physical phenomena of the above control operations are as follows. At 2.0 s, the actuator thrust direction undergoes a rapid change from negative to positive. As shown in Fig. 4, due to the presence of backlash clearance in the powertrain, the dynamics transition from the contact mode to the backlash mode and then back to the contact mode. During the backlash mode, the dead-zone characteristic prevents the actuator from transmitting force to the vehicle body. Since MBC does not explicitly account for these dynamics, it computes the excessively large control input during the backlash mode. With only MBC, a shock will be induced at the moment of transition back to the contact mode. In contrast, RL serves to correct this error. Specifically, RL applies the control input in the opposite direction to MBC, providing an appropriate braking effect to the actuator. This achieves a "soft-landing" effect, effectively reducing the shock.

Another noteworthy aspect of this study is the achievement of superior performance by the proposed method, despite utilizing limited data and a relatively simple neural network architecture. As shown in Table 2, the number of training episodes in this study is 300, which is significantly lower than that of similar applications to mechanical system control with domain randomization in related research [32]. Furthermore, the required number of neurons is only 64. The generalization ability to randomized environments achieved under such simplified conditions is attributed to the introduction of MBC. In the proposed method, the RL agent does not need to explore the adequate control action from scratch, because MBC provides foundational knowledge regarding how to control system dynamics. During the learning process, therefore, the RL agent primarily focuses on enhancing robustness, which is considered less complex than constructing a base for control action independently. To facilitate this, the proposed method shares the knowledge on control input from MBC with the RL agent as a learning cue. To further confirm whether the proposed method can overcome the sim-to-real gap, Fig. 19 shows the results of a statistical evaluation of the control system. In this verification, 100 different conditions were generated by randomly varying the parameters of the powertrain, and statistical analysis was performed for all of these patterns. Fig. 19 presents the time-history average across the 100 verification cases. Furthermore, Table 13 summarizes the mean and standard deviation of the 2-norm values of the responses. As can be seen, the proposed method exhibited the highest robustness against 100 powertrain variations that differed from the training environment. From these results, it is demonstrated that the proposed method achieves sim-to-real transfer in verifications designed to emulate actual systems where arbitrary parameter variations may occur.

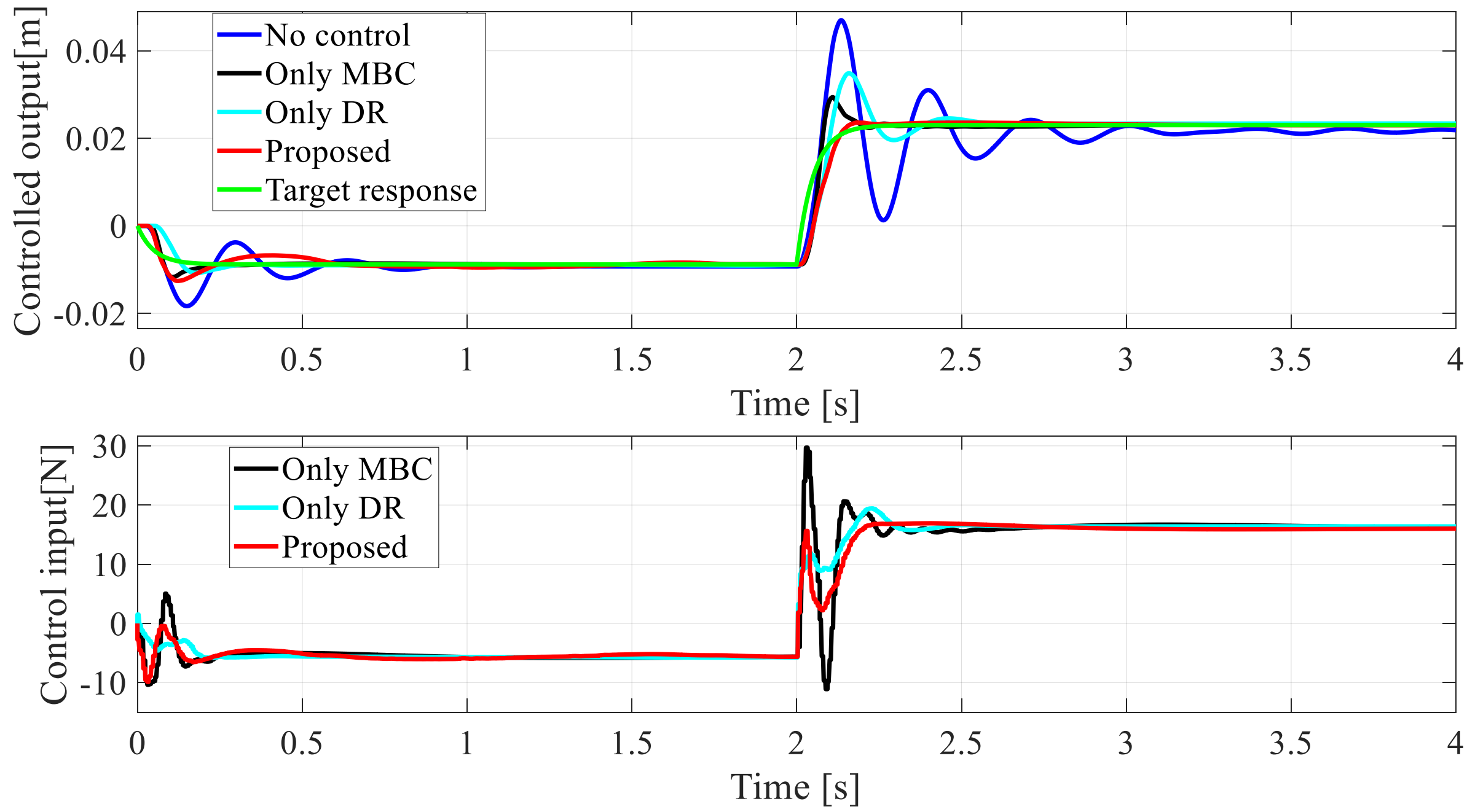

Fig. 19 Mean time responses of 100 trials when random variations are applied to the powertrain parameters in *Scenario 2*.

Table 13 Statistics of 2-norm of 100 trials computed for the control results in Fig. 19.

| | Proposed method (MBCA-DRT) | Only domain randomization | Only MBC | Open-loop (i.e., no control) |
|---|---|---|---|---|
| Mean values of 2-norm | 0.9214 | 1.3097 | 0.9263 | 3.5179 |
| Standard deviation | 0.1584 | 0.2255 | 0.1707 | 0.7346 |

### 5.5. Verification on robustness against additional possible uncertainty

This section investigates the robustness of the proposed method against uncertainties other than parameter variations in the controlled plant. In an actual powertrain system, observation noise inevitably affects sensor measurements. To reflect this, Gaussian noise with a standard deviation of 5% of the observation value was added in the training environment. Furthermore, to account for the vehicle dynamics such as resistance and friction acting on a vehicle body, a frictional force model was incorporated. Figures 20 and 21 show the control performance obtained when these uncertainties were considered in the training environment. For example, the magenta line shows the open-loop response when observation noise was applied to the vehicle body vibration. As shown, the proposed method consistently achieved effective vibration suppression even when the powertrain includes observation noise and friction, demonstrating its robustness. Tables 14 and 15 provide quantitative performance comparisons, further confirming the validity of the proposed approach. These results demonstrate that incorporating sensor noise and friction into the training environment enhances the robustness provided by domain randomization and improves the generalization capability of the actor network.

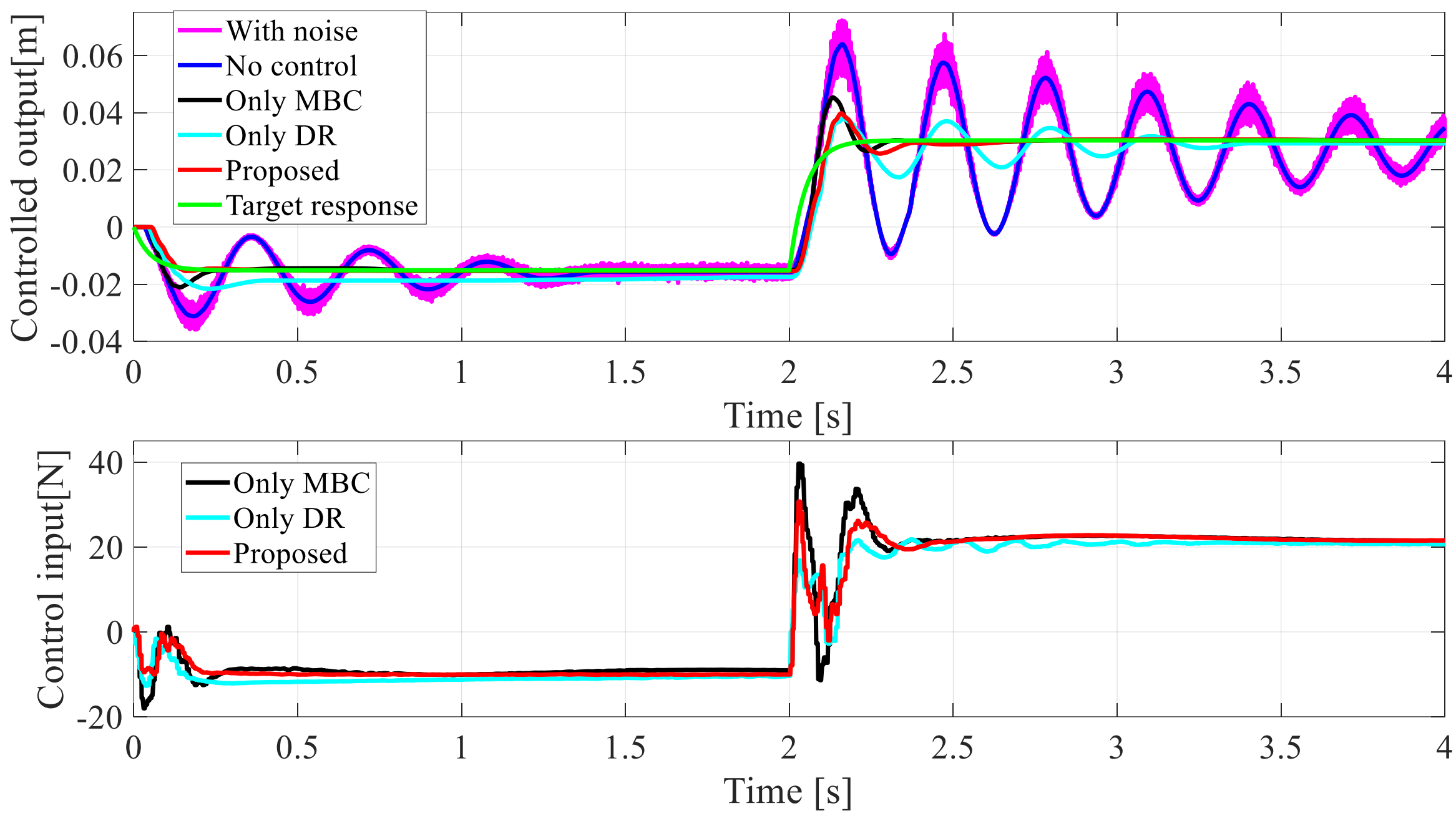

Fig. 20 Control result when the powertrain includes noise and friction with $M_B = M_B^{max}$, $M_E = M_E^{max}$, $y_{0-2s}^r = y_{0-2s}^{r\,min}$, $y_{2-4s}^r = y_{2-4s}^{r\,max}$ and $\delta = \delta_{min}$.

Table 14 2-norm computed for the control results in Fig. 20.

| | Proposed method (MBCA-DRT) | Only domain randomization | Only MBC | Open-loop (i.e., no control) |
|---|---|---|---|---|
| 2-norm | 1.3226 | 2.1658 | 1.4508 | 5.6867 |

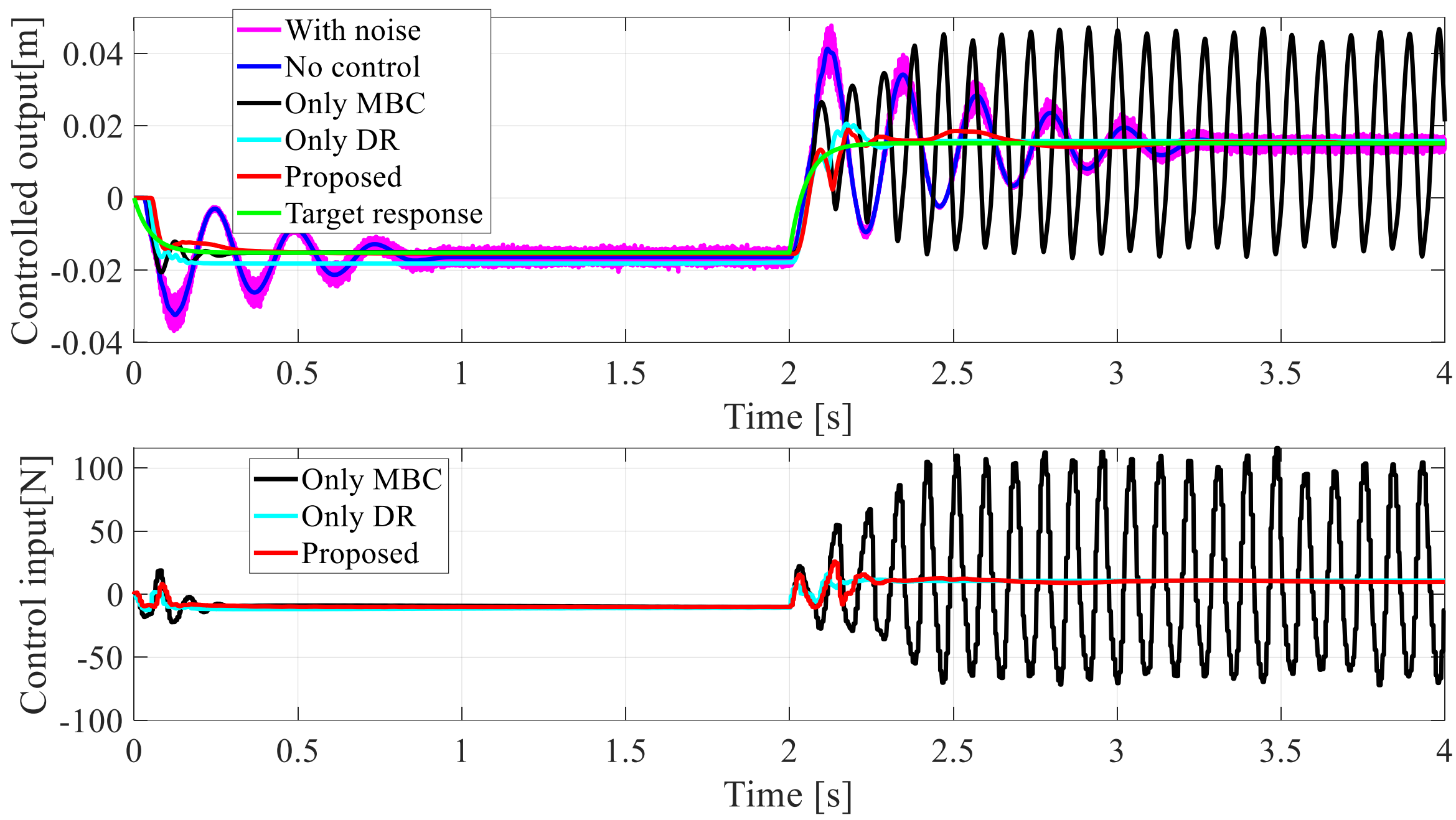

Fig. 21 Control result when the powertrain includes noise and friction with $M_B = M_B^{max}$, $M_E = M_E^{min}$, $y_{0-2s}^r = y_{0-2s}^{r\,min}$, $y_{2-4s}^r = y_{2-4s}^{r\,min}$ and $\delta = \delta_{max}$.

Table 15 2-norm computed for the control results in Fig. 21.

| | Proposed method (MBCA-DRT) | Only domain randomization | Only MBC | Open-loop (i.e., no control) |
|---|---|---|---|---|
| 2-norm | 0.8412 | 1.1782 | 6.1651 | 3.1198 |

The central claim of this study is that we have demonstrated *the potential for achieving powerful robustness of control even with limited data by integrating MBC and RL*. From a broader perspective, this effectiveness is attributable to the *physics (e.g., first principles model)-based reliability* of MBC. Even in the current era of advanced AI, "lazy" approaches that solely rely on black-box machine learning for rudely replacing all existing control systems present inherent limitations. Consequently, the role of MBC is expected to remain significant in future developments.

## 6. Conclusion

For mechanical systems subject to nonlinearities and uncertainties, this study presents a novel robust control method based on DRL. The proposed approach relies on the effective combination of domain randomization, LSTM-based actor and critic networks, and utilization of MBC. We develop a robust DDPG algorithm where learning progress is boosted by the hybridized control of RL and MBC, preventing the resultant policy from being too conservative. The proposed approach is verified for active vibration control of a complex powertrain system with backlash nonlinearity and parametric uncertainty. Compared to conventional control approaches, numerical examples confirm that the proposed approach provides excellent damping performance and stronger robustness against system variations.

In the future, experimental verifications will be conducted by implementing the proposed control system in a basic experimental device reflecting a simplified powertrain system. In addition, future work will extend the verification to more complex domain randomization involving a larger combination of physical parameters.

### CRediT authorship contribution statement

**Heisei Yonezawa**: Conceptualization, Methodology, Software, Validation, Investigation, Writing - Original Draft. **Ansei Yonezawa**: Conceptualization, Validation, Writing - Original Draft. **Itsuro Kajiwara**: Conceptualization, Writing - Original Draft, Supervision.

### Declaration of competing interest

The authors declare that they have no known competing financial interests or personal relationships that could have appeared to influence the work reported in this paper.

### Acknowledgment

A part of this work was supported by the Japan Society for the Promotion of Science (JSPS) KAKENHI [Grant Number 23K13273].

## References

[1] F. Wang, T. Wu, Y. Ni, P. Ye, Y. Cai, J. Guo, C. Wang, Torsional oscillation suppression-oriented torque compensate control for regenerative braking of electric powertrain based on mixed logic dynamic model, Mech. Syst. Signal Process. 190 (2023) 110114. https://doi.org/10.1016/j.ymssp.2023.110114.

[2] F. Wang, P. Ye, X. Xu, Y. Cai, S. Ni, H. Que, Novel regenerative braking method for transient torsional oscillation suppression of planetary-gear electrical powertrain, Mech. Syst. Signal Process. 163 (2022) 108187. https://doi.org/10.1016/j.ymssp.2021.108187.

[3] C. Wu, K. Guo, J. Sun, Y. Liu, D. Zheng, Active vibration control in robotic grinding using six-axis acceleration feedback, Mech. Syst. Signal Process. 214 (2024) 111379. https://doi.org/10.1016/j.ymssp.2024.111379.

[4] S. Zaare, M.R. Soltanpour, Adaptive fuzzy global coupled nonsingular fast terminal sliding mode control of n-rigid-link elastic-joint robot manipulators in presence of uncertainties, Mech. Syst. Signal Process. 163 (2022) 108165. https://doi.org/10.1016/j.ymssp.2021.108165.

[5] Y. Yang, C. Liu, T. Zhang, X. Zhou, J. Li, Event-triggered composite sliding mode anti-sway control for tower crane systems and experimental verification, Mech. Syst. Signal Process. 230 (2025) 112578. https://doi.org/10.1016/j.ymssp.2025.112578.

[6] Z. Xu, G. Qi, Q. Liu, J. Yao, ESO-based adaptive full state constraint control of uncertain systems and its application to hydraulic servo systems, Mech. Syst. Signal Process. 167 (2022) 108560. https://doi.org/10.1016/j.ymssp.2021.108560.

[7] R.E. Kalman, On the general theory of control systems, IFAC Proc. Vol. 1 (1960) 491–502. https://doi.org/10.1016/S1474-6670(17)70094-8.

[8] X. Bin Peng, M. Andrychowicz, W. Zaremba, P. Abbeel, Sim-to-Real Transfer of Robotic Control with Dynamics Randomization, in: 2018 IEEE Int. Conf. Robot. Autom., IEEE, 2018: pp. 3803–3810. https://doi.org/10.1109/ICRA.2018.8460528.

[9] S. Xu, X. Liu, Y. Wang, Z. Sun, J. Wu, Y. Shi, Frequency shaping-based H∞ control for active pneumatic vibration isolation with input voltage saturation, Mech. Syst. Signal Process. 220 (2024) 111705. https://doi.org/10.1016/j.ymssp.2024.111705.

[10] A. Noormohammadi-Asl, O. Esrafilian, M. Ahangar Arzati, H.D. Taghirad, System identification and H∞-based control of quadrotor attitude, Mech. Syst. Signal Process. 135 (2020) 106358. https://doi.org/10.1016/j.ymssp.2019.106358.

[11] Y. Zhao, Y. Zhang, L. Guo, S. Ding, X. Wang, Advances in machine learning-based active vibration control for automotive seat suspensions: A comprehensive review, Mech. Syst. Signal Process. 231 (2025) 112645. https://doi.org/10.1016/j.ymssp.2025.112645.

[12] Z. Qiu, Z. Hu, Biological evolution reinforcement learning vibration control of a three-flexible-beam coupling multi-body system, Mech. Syst. Signal Process. 231 (2025) 112634.

https://doi.org/10.1016/j.ymssp.2025.112634.
[13] D. Chen, P. Yu, G. Wang, X. Liu, Y. Ding, J. Jin, Design of a hybrid-mode piezoelectric actuator for compact robotic finger based on deep reinforcement learning, Mech. Syst. Signal Process. 227 (2025) 112401. https://doi.org/10.1016/j.ymssp.2025.112401.
[14] R.S. Sutton, A.G. Barto, Reinforcement learning: An introduction, MIT Press, Cambridge, 1998.
[15] R.S. Sutton, A.G. Barto, Reinforcement learning: An introduction, 2nd ed., MIT Press, Cambridge, 2018.
[16] V. Mnih, K. Kavukcuoglu, D. Silver, A. Graves, I. Antonoglou, D. Wierstra, M. Riedmiller, Playing Atari with Deep Reinforcement Learning, ArXiv Prepr. (2013). https://doi.org/https://doi.org/10.48550/arXiv.1312.5602.
[17] V. Mnih, K. Kavukcuoglu, D. Silver, A.A. Rusu, J. Veness, M.G. Bellemare, A. Graves, M. Riedmiller, A.K. Fidjeland, G. Ostrovski, S. Petersen, C. Beattie, A. Sadik, I. Antonoglou, H. King, D. Kumaran, D. Wierstra, S. Legg, D. Hassabis, Human-level control through deep reinforcement learning, Nature. 518 (2015) 529–533. https://doi.org/10.1038/nature14236.
[18] D. Silver, G. Lever, N. Heess, T. Degris, D. Wierstra, M. Riedmiller, Deterministic policy gradient algorithms, in: 31st Int. Conf. Mach. Learn. ICML 2014, 2014: pp. 387–395.
[19] T.P. Lillicrap, J.J. Hunt, A. Pritzel, N. Heess, T. Erez, Y. Tassa, D. Silver, D. Wierstra, Continuous control with deep reinforcement learning, in: 4th Int. Conf. Learn. Represent. ICLR 2016 - Conf. Track Proc., 2015. http://arxiv.org/abs/1509.02971.
[20] S. Fujimoto, H. Van Hoof, D. Meger, Addressing Function Approximation Error in Actor-Critic Methods, in: 35th Int. Conf. Mach. Learn. PMLR, 2018: pp. 1587–1596. https://doi.org/https://doi.org/10.48550/arXiv.1802.09477.
[21] J. Schulman, F. Wolski, P. Dhariwal, A. Radford, O. Klimov, Proximal Policy Optimization Algorithms, ArXiv Prepr. (2017). https://doi.org/https://doi.org/10.48550/arXiv.1707.06347.
[22] L. Ouyang, J. Wu, X. Jiang, D. Almeida, C.L. Wainwright, P. Mishkin, C. Zhang, S. Agarwal, K. Slama, A. Ray, J. Schulman, J. Hilton, F. Kelton, L. Miller, M. Simens, A. Askell, P. Welinder, P. Christiano, J. Leike, R. Lowe, Training language models to follow instructions with human feedback, Adv. Neural Inf. Process. Syst. 35 (2022) 27730–27744. http://arxiv.org/abs/2203.02155.
[23] R. Zheng, S. Dou, S. Gao, Y. Hua, W. Shen, B. Wang, Y. Liu, S. Jin, Q. Liu, Y. Zhou, L. Xiong, L. Chen, Z. Xi, N. Xu, W. Lai, M. Zhu, C. Chang, Z. Yin, R. Weng, W. Cheng, H. Huang, T. Sun, H. Yan, T. Gui, Q. Zhang, X. Qiu, X. Huang, Secrets of RLHF in Large Language Models Part I: PPO, ArXiv Prepr. (2023). https://doi.org/https://doi.org/10.48550/arXiv.2307.04964.
[24] J.K. Viswanadhapalli, V.K. Elumalai, S. S., S. Shah, D. Mahajan, Deep reinforcement learning with reward shaping for tracking control and vibration suppression of flexible link manipulator, Appl. Soft Comput. 152 (2024) 110756. https://doi.org/10.1016/j.asoc.2023.110756.
[25] M.H. Ahmed, A. AboHussien, A. El-Shafei, A.M. Darwish, A.H. Abdel-Gawad, Active control of flexible rotors using deep reinforcement learning with application of multi-actor-critic deep

deterministic policy gradient, Eng. Appl. Artif. Intell. 124 (2023) 106593. https://doi.org/10.1016/j.engappai.2023.106593.
[26] C. Wang, W. Cheng, H. Zhang, W. Dou, J. Chen, An immune optimization deep reinforcement learning control method used for magnetorheological elastomer vibration absorber, Eng. Appl. Artif. Intell. 137 (2024) 109108. https://doi.org/10.1016/j.engappai.2024.109108.
[27] Z. Qiu, Y. Liu, X. Zhang, Reinforcement learning vibration control and trajectory planning optimization of translational flexible hinged plate system, Eng. Appl. Artif. Intell. 133 (2024) 108630. https://doi.org/10.1016/j.engappai.2024.108630.
[28] Y.-A. Zhang, S. Zhu, Novel Model-free Optimal Active Vibration Control Strategy Based on Deep Reinforcement Learning, Struct. Control Heal. Monit. 2023 (2023) 1–15. https://doi.org/10.1155/2023/6770137.
[29] J. Panda, M. Chopra, V. Matsagar, S. Chakraborty, Continuous control of structural vibrations using hybrid deep reinforcement learning policy, Expert Syst. Appl. 252 (2024) 124075. https://doi.org/10.1016/j.eswa.2024.124075.
[30] G. Chen, Z. Chen, L. Wang, W. Zhang, Deep Deterministic Policy Gradient and Active Disturbance Rejection Controller based coordinated control for gearshift manipulator of driving robot, Eng. Appl. Artif. Intell. 117 (2023) 105586. https://doi.org/10.1016/j.engappai.2022.105586.
[31] T. Slawik, B. Wehbe, L. Christensen, F. Kirchner, Deep Reinforcement Learning for Path-Following Control of an Autonomous Surface Vehicle using Domain Randomization, IFAC-PapersOnLine. 58 (2024) 21–26. https://doi.org/10.1016/j.ifacol.2024.10.027.
[32] J. Zhang, C. Zhao, J. Ding, Deep reinforcement learning with domain randomization for overhead crane control with payload mass variations, Control Eng. Pract. 141 (2023) 105689. https://doi.org/10.1016/j.conengprac.2023.105689.
[33] X. Chen, J. Hu, C. Jin, L. Li, L. Wang, Understanding Domain Randomization for Sim-to-real Transfer, in: ICLR 2022 - 10th Int. Conf. Learn. Represent., 2021: pp. 1–28. https://doi.org/https://doi.org/10.48550/arXiv.2110.03239.
[34] J. Matas, S. James, A.J. Davison, Sim-to-Real Reinforcement Learning for Deformable Object Manipulation, in: 2nd Conf. Robot Learn. PMLR, 2018: pp. 734–743. http://arxiv.org/abs/1806.07851.
[35] Z. Li, X. Cheng, X. Bin Peng, P. Abbeel, S. Levine, G. Berseth, K. Sreenath, Reinforcement Learning for Robust Parameterized Locomotion Control of Bipedal Robots, in: 2021 IEEE Int. Conf. Robot. Autom., IEEE, 2021: pp. 2811–2817. https://doi.org/10.1109/ICRA48506.2021.9560769.
[36] X. Bin Peng, E. Coumans, T. Zhang, T.-W. Lee, J. Tan, S. Levine, Learning Agile Robotic Locomotion Skills by Imitating Animals, in: Robot. Sci. Syst., Robotics: Science and Systems Foundation, 2020. https://doi.org/10.15607/RSS.2020.XVI.064.
[37] Q. Liao, B. Zhang, X. Huang, X. Huang, Z. Li, K. Sreenath, Berkeley Humanoid: A Research

Platform for Learning-based Control, ArXiv Prepr. (2024). https://doi.org/https://doi.org/10.48550/arXiv.2407.21781.

[38] X. Gu, Y.-J. Wang, X. Zhu, C. Shi, Y. Guo, Y. Liu, J. Chen, Advancing Humanoid Locomotion: Mastering Challenging Terrains with Denoising World Model Learning, in: Robot. Sci. Syst., Robotics: Science and Systems Foundation, 2024. https://doi.org/10.15607/RSS.2024.XX.058.

[39] A. Loquercio, E. Kaufmann, R. Ranftl, A. Dosovitskiy, V. Koltun, D. Scaramuzza, Deep Drone Racing: From Simulation to Reality With Domain Randomization, IEEE Trans. Robot. 36 (2020) 1–14. https://doi.org/10.1109/TRO.2019.2942989.

[40] O. Nachum, M. Ahn, H. Ponte, S. Gu, V. Kumar, Multi-Agent Manipulation via Locomotion using Hierarchical Sim2Real, Proc. Mach. Learn. Res. 100 (2019) 110–121. https://doi.org/https://doi.org/10.48550/arXiv.1908.05224.

[41] F. Ramos, R. Possas, D. Fox, BayesSim: Adaptive Domain Randomization Via Probabilistic Inference for Robotics Simulators, in: Robot. Sci. Syst. XV, Robotics: Science and Systems Foundation, 2019. https://doi.org/10.15607/RSS.2019.XV.029.

[42] Y. Cheng, P. Zhao, F. Wang, D.J. Block, N. Hovakimyan, Improving the Robustness of Reinforcement Learning Policies With L1 Adaptive Control, IEEE Robot. Autom. Lett. 7 (2022) 6574–6581. https://doi.org/10.1109/LRA.2022.3169309.

[43] Z. Yao, X. Liang, S. Wang, J. Yao, Model-Data Hybrid Driven Control of Hydraulic Euler–Lagrange Systems, IEEE/ASME Trans. Mechatronics. 30 (2025) 131–143. https://doi.org/10.1109/TMECH.2024.3390129.

[44] L.P. Kaelbling, M.L. Littman, A.W. Moore, Reinforcement Learning: A Survey, J. Artif. Intell. Res. 4 (1996) 237–285. https://doi.org/10.1613/jair.301.

[45] R.S. Sutton, D. McAllester, S. Singh, Y. Mansour, Policy gradient methods for reinforcement learning with function approximation, in: Adv. Neural Inf. Process. Syst. 12, 2000: pp. 1057–1063.

[46] J. Peters, S. Vijayakumar, S. Schaal, Natural Actor-Critic, in: 16th Eur. Conf. Mach. Learn., 2005: pp. 280–291. https://doi.org/10.1007/11564096_29.

[47] S. Bhatnagar, R.S. Sutton, M. Ghavamzadeh, M. Lee, Incremental natural actor-critic algorithms, in: Adv. Neural Inf. Process. Syst. 20 - Proc. 2007 Conf., 2007.

[48] T. Degris, P.M. Pilarski, R.S. Sutton, Model-Free reinforcement learning with continuous action in practice, in: 2012 Am. Control Conf., IEEE, 2012: pp. 2177–2182. https://doi.org/10.1109/ACC.2012.6315022.

[49] T. Degris, M. White, R.S. Sutton, Off-Policy Actor-Critic, in: Proc. 29th Int. Conf. Mach. Learn. ICML 2012, 2012: pp. 457–464. http://arxiv.org/abs/1205.4839.

[50] C.J.C.H. Watkins, P. Dayan, Q-learning, Mach. Learn. 8 (1992) 279–292. https://doi.org/10.1007/BF00992698.

[51] J. Kwon, Y. Efroni, C. Caramanis, S. Mannor, RL for Latent MDPs: Regret Guarantees and a Lower Bound, Adv. Neural Inf. Process. Syst. 34 (2021) 24523–24534.

https://doi.org/https://doi.org/10.48550/arXiv.2102.04939.

[52] Y. Okawa, T. Sasaki, H. Iwane, Control Approach Combining Reinforcement Learning and Model-Based Control, in: 2019 12th Asian Control Conf. ASCC 2019, JSME, 2019: pp. 1419–1424.

[53] I. Koryakovskiy, M. Kudruss, H. Vallery, R. Babuska, W. Caarls, Model-Plant Mismatch Compensation Using Reinforcement Learning, IEEE Robot. Autom. Lett. 3 (2018) 2471–2477. https://doi.org/10.1109/LRA.2018.2800106.

[54] S. Hochreiter, J. Schmidhuber, Long Short-Term Memory, Neural Comput. 9 (1997) 1735–1780. https://doi.org/10.1162/neco.1997.9.8.1735.

[55] H. Yonezawa, A. Yonezawa, T. Hatano, S. Hiramatsu, C. Nishidome, I. Kajiwara, Fuzzy-reasoning-based robust vibration controller for drivetrain mechanism with various control input updating timings, Mech. Mach. Theory. 175 (2022) 104957. https://doi.org/10.1016/j.mechmachtheory.2022.104957.

[56] H. Yonezawa, A. Yonezawa, I. Kajiwara, Experimental validation of adaptive grey wolf optimizer-based powertrain vibration control with backlash handling, Mech. Mach. Theory. 203 (2024) 105825. https://doi.org/10.1016/j.mechmachtheory.2024.105825.

[57] M. Chilali, P. Gahinet, H∞ design with pole placement constraints: An LMI approach, IEEE Trans. Automat. Contr. 41 (1996) 358–367. https://doi.org/10.1109/9.486637.

[58] Zhou K, Doyle JC, Glover K, Robust and Optimal Control, PrenticeHall, New Jersey, 1996.

[59] H. Yonezawa, A. Yonezawa, I. Kajiwara, Grey wolf optimization tuned drivetrain vibration controller with backlash compensation strategy using time-dependent-switched Kalman filter, Proc. Inst. Mech. Eng. Part D J. Automob. Eng. (2024). https://doi.org/10.1177/09544070241240019.

[60] R. Liessner, J. Schmitt, A. Dietermann, B. Bäker, Hyperparameter Optimization for Deep Reinforcement Learning in Vehicle Energy Management, in: Proc. 11th Int. Conf. Agents Artif. Intell., SCITEPRESS - Science and Technology Publications, 2019: pp. 134–144. https://doi.org/10.5220/0007364701340144.